\documentclass{pasj00}
\newcounter{draftenable}
\begin{document}

\SetRunningHead{Accretion Geometry of the LMXB Aql X-1}{Sakurai et al.}

\title{Accretion Geometry of the Low-Mass X-ray Binary Aquila X-1 in the Soft and Hard States}


\author{Soki \textsc{Sakurai}, Shin'ya \textsc{Yamada}, Shunsuke \textsc{Torii}, Hirofumi \textsc{Noda}, Kazuhiro \textsc{Nakazawa}, and Kazuo \textsc{Makishima}
 }
\affil{Department of Physics, University of Tokyo 7-3-1, Hongo, Bonkyo-ward, Tokyo, 113-0033}
\email{sakurai@juno.phys.s.u-tokyo.ac.jp}

\and
\author{Hiromitsu \textsc{Takahashi}}
\affil{High Energy Astrophysics Group, Department of Physical Sciences
Hiroshima University
1-3-1 Kagamiyama, Higashi-Hiroshima, Hiroshima, 739-8526}



%


\maketitle

\begin{abstract}
The neutron-star Low-Mass X-ray Binary Aquila X-1 was observed seven times in total with the Suzaku X-ray observatory from September 28 to October 30 in 2007, in the decaying phase of an outburst.  
  In order to constrain the flux-dependent accretion geometry of this source over wider energy bands than employed in most of previous works, the present study utilized two out of the seven data sets.
  The 0.8--31 keV spectrum on September 28, taken with the XIS and HXD-PIN for an exposure of 13.8 ks, shows an absorbed 0.8--31 keV flux of $3.6\times 10^{-9}$ erg s$^{-1}$ cm$^{-2}$, together with typical characteristics of the soft state of this type of objects.
  The spectrum was successfully explained by 
an optically-thick disk emission plus a Comptonized blackbody component.
Although these results are in general agreement with previous studies, 
the significance of a hard tail recently reported using the same data 
was inconclusive in our analysis.
The spectrum acquired on October 9 for an exposure of 19.7 ks was detected over a 0.8--100 keV band with the XIS, HXD-PIN, and HXD-GSO, at an absorbed flux of $8.5\times 10^{-10}$ erg s$^{-1}$ cm$^{-2}$ (in 0.8--100 keV).  It shows characteristics of the hard state, and was successfully explained by the same two continuum components but with rather different parameters including much stronger thermal Comptonization, of which the seed photon source was identified with blackbody emission from the neutron-star surface.
  As a result, the accretion flow in the hard state is inferred to take a form of an optically-thick and geometrically-thin disk down to a radius of $21\pm 4$ km from the neutron star, and then turn into an optically-thin nearly-spherical hot flow.
\end{abstract}

\section{Introduction}
\begin{figure}[htbp]
  \begin{center}
    \FigureFile(80mm,80mm){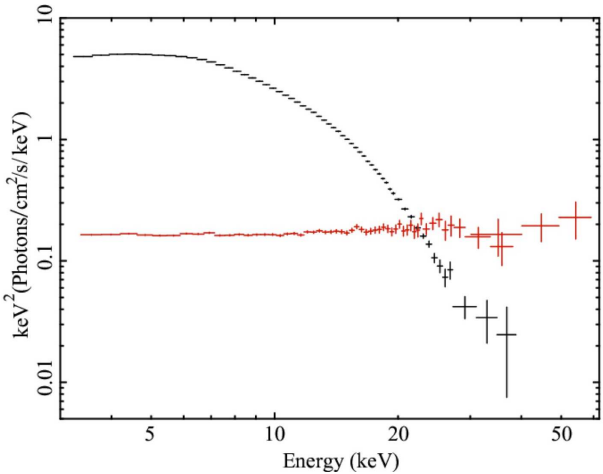}
  \end{center}
  \caption{Background-subtracted $\nu F\nu$ Spectra of Aquila X-1 in the soft state (black) and hard state (red), derived by analyzing archival RXTE PCA and HEXTE data obtained on 1998 March 12 and 1999 September 2, respectively.}\label{fig:intro_softandhard}
\end{figure}

A neutron-star Low-Mass X-ray Binary (NS-LMXB) consists of a weakly magnetized neutron star and a low-mass star with the mass typically less than that of the Sun, $M_\odot$.  NS-LMXBs have been studied from the 1980's and were found to dominate the X-ray luminosities of galaxies without strong Active Galactic Nuclei \citep{Makishima1989}.  
Like Black Hole Binaries (BHBs), NS-LMXBs are known to exhibit soft and hard spectral states, when their mass accretion rate is high and low, respectively \citep{White1985,Mitsuda1986,Mitsuda1989}.    The two types of spectra are visualized in figure \ref{fig:intro_softandhard}; it compares a soft- and hard-state spectrum of the NS-LMXB Aquila X-1 which we derived by analyzing archival RXTE PCA and HEXTE data.

In the soft state, the spectrum is bright  in a soft band below 10--15 keV, and falls off steeply towards higher energies.  It was shown with Tenma that  the soft-state spectra of NS-LMXBs can be expressed by the sum of a blackbody component from the NS surface and a disk blackbody component \citep{Mitsuda1984};
 the latter is defined as a particular superposition of blackbodies with a range of temperature, and is considered to approximate the integrated emission from an optically-thick standard accretion disk \citep{Shakura_Sunyaev1973}.  The validity of this model composition has been confirmed repeatedly by subsequent studies \citep[e.g.][]{Makishima1989,Asai1998,Lin2010,Takahashi2011}.
  For comparison, spectra of BHBs in the soft state are even softer \citep{White1984,Makishima1986}, because black holes lack a solid surface which would emit the relatively hard blackbody component.

  As represented by figure \ref{fig:intro_softandhard}, spectra of NS-LMXBs in the hard state are roughly power-law-like with a photon index of $\Gamma\sim 2$, and often extend up to 100 keV \citep[e.g.][]{Tarana2006}.
  Hence in the hard state, broad-band spectroscopy becomes more essential in qualifying major emission components.
  In addition, the luminosity of an NS-LMXB in the hard state is not much higher than $\sim10^{36} $ erg s$^{-1}$.
  As a result, the study of NS-LMXBs in the hard state is often limited by the available hard X-ray instrumentation, and has been making a considerable progress recently thanks to high hard X-ray sensitivities of BeppoSax, RXTE, INTEGRAL, and Swift.
It has been reported that the spectra can be expressed by the sum of an 
optically thick emission in the soft band and a Comptonized emission spanning nearly the whole band \citep{Lin2007,Cocchi2011,Tarana2011}.
  Also, detailed timing studies are revealing differences between NS-LMXBs and BHBs focusing on their hard X-ray properties \citep{Titarchuk2005}.
  However, we still need to uniquely associate the spectral emission components to actual accretion flows, including the location of the optically-thick emitter, the source supplying seed photons of Comptonization, and the geometry of the Compton corona.
  Since Suzaku \citep{Mitsuda2007} allows us to acquire the spectra of hard-state LMXBs over a very wide (from $\leq 1$ keV to $\geq 100$ keV) energy band and hence to simultaneously quantify the proposed emission components, we expect to achieve significant progresses on these issues \citep[e.g.][]{Lin2010}.
 
  In the present research, we focus on the NS-LMXB Aquila X-1, and investigate its soft and hard states in detail utilizing archived Suzaku data.  Aquila X-1 is a recurrent transient with a  2--10 keV luminosity varying by a factor of $>$ 200 through its outbursts.
  It is also an emitter of Type I X-ray bursts \citep{Koyama1981}, and its distance is estimated as 4.4--5.9 kpc by assuming that the peak luminosity of its Type I X-ray bursts equals to the Eddington luminosity for a $1.4M_\odot$ neutron star \citep{Jonker2004}.  It was observed by Suzaku seven times in 2007, all covering a declining phase of a single outburst.
  Although these data sets were already utilized by \cite{Raichur2011}, hereafter RMD11, they did not use the HXD-GSO data even for the hard state, and hence the available information was limited to $\leq 70$ keV with relatively large errors above $\sim 50$ keV.  We hence try utilizing the GSO data for the hard state, taking into account the latest calibration results \citep{Yamada2011}.

\section{Observation and Data Processing}
\begin{figure}[htbp]
  \begin{center}
    \FigureFile(80mm,80mm){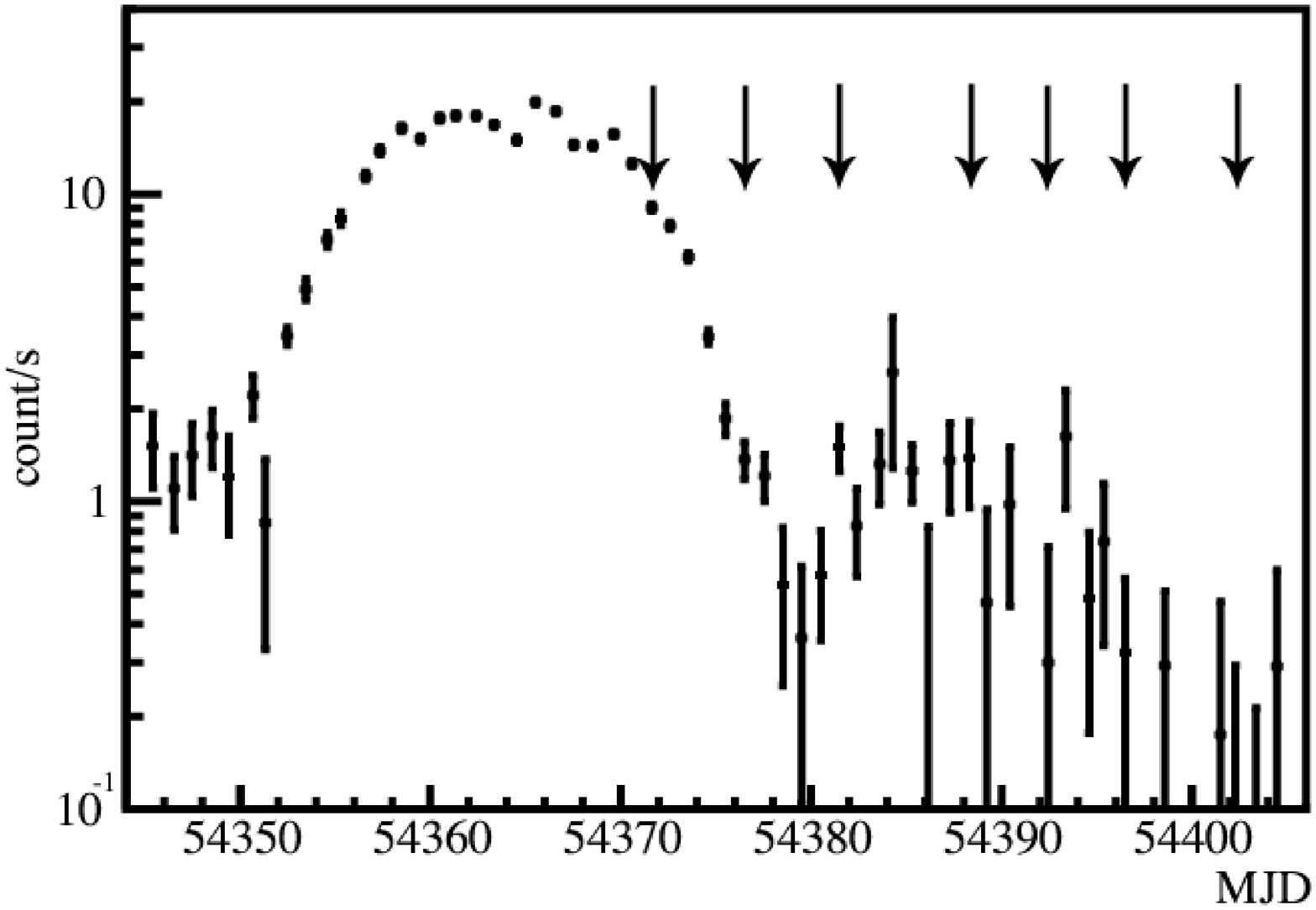}
  \end{center}
  \caption{A soft band (1.5--12 keV) light curve of Aquila X-1 from 2007 September 1 to 2007 October 31 obtained with the RXTE ASM.  Arrows indicate the 7 Suzaku observations.}\label{fig:LCasm}
\end{figure}

Aquila X-1 was observed with Suzaku from September 28 to October 30 in 2007 for seven times.
  The log of these observations is shown in table \ref{tb:Obss}.
  Figure \ref{fig:LCasm} is a soft-band (1.5--12 keV) light curve obtained with the RXTE ASM.  Of these 7 observations, RMD11 analyzed the first four.

\begin{table*}[htbp]
\caption{Suzaku observations of Aql X-1 in 2007.}
\centering
\begin{minipage}{11cm}
\begin{tabular}{cccccc}\hline
No.&Date\footnotemark[$*$]  & ObsID & Exp\footnotemark[$\dagger$] & \multicolumn{2}{c}{Count rate\footnotemark[$\ddagger$]} \\\cline{5-6}
&&& (ks) & XIS  & HXD-PIN \\\hline
1&9/28 & 402053010 & 13.8 & 275.4$\pm$0.4\footnotemark[$\#$]  & 1.12$\pm$0.01 \\
2&10/03 &402053020&15.1  & 28.12$\pm$0.04 & 0.74$\pm$0.01 \\
3&10/09 &402053030&19.7  & 33.18$\pm$0.04 & 0.89$\pm$0.01\\
4&10/15 &402053040&17.9  & 25.85$\pm$0.04 & 0.73$\pm$0.01 \\
5&10/19 &402053050&17.9  & 5.67$\pm$0.02 & 0.13$\pm$0.01 \\
6&10/24 &402053060&21.4  & 0.221$\pm$0.003 & 0.041$\pm$0.005 \\
7&10/30 &402053070&17.5  & 0.172$\pm$0.003 & 0.056$\pm$0.006 \\\hline
\end{tabular}
\label{tb:Obss}
\footnotetext[$*$]{Date in 2007.}
\footnotetext[$\dagger$]{Net exposure per XIS sensor.}
\footnotetext[$\ddagger$]{In ct s$^{-1}$ after subtracting the background.  The XIS rate is in 0.8--10 keV, with the two FI sensors (XIS0, XIS3) summed.   The HXD-PIN rate refers to a 12--50 keV range, including the CXB contribution of $\sim 0.02$ ct s$^{-1}$.  The quoted errors are statistical 1$\sigma$ limits.}
\footnotetext[$\#$]{Corrected for pile-up effects (see text).}
\end{minipage}
\end{table*}%

\begin{figure}[htbp]
  \begin{center}
    \FigureFile(80mm,80mm){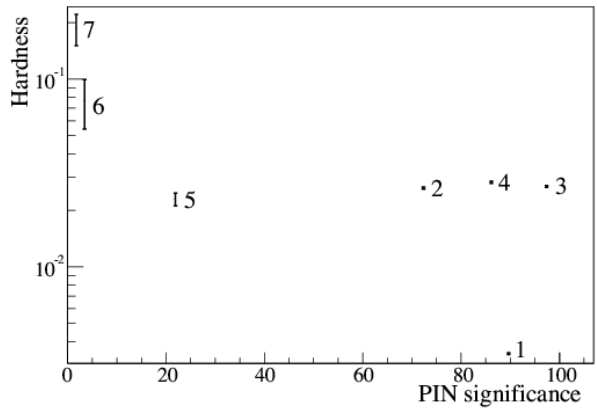}
  \end{center}
  \caption{Characterization of the 7 Suzaku observations in terms of PIN significance and the signal hardness, the latter defined by the HXD-PIN vs XIS count-rate ratio.  See text (subsection \ref{ss:DP}) for the definition of the two quantities.  Vertical error bars are statistical only.}\label{fig:HNR}
\end{figure}

\subsection{Characterization of the 7 data sets}\label{ss:DP}
By quickly analyzing the seven Suzaku datasets processed through HEADAS (version 6.9), we obtained the source count rates as given in table \ref{tb:Obss}.  The processing details are described in subsection \ref{ss:DataProcessing}.
The derived XIS count rates for the first four data sets agree with those in RMD11, except for a factor 2 difference because they refer to a single XIS sensor (though not explicitly stated) while we employ the XIS0 plus XIS3 rates.
  Our HXD-PIN rate is comparable too, but systematically lower by $\sim 0.03$ ct s$^{-1}$, particularly in the 2nd--4th observations, than those of RMD11: this is presumably because they used a different (non specified) energy band than the 12--50 keV band which we used.

  To grasp the source characteristics in the seven observations, we calculated, for each of them, spectral hardness and HXD-PIN significance.  The former is defined as the HXD-PIN vs XIS count-rate ratio referring to table \ref{tb:Obss}; the latter is given as the background-subtracted 12--50 keV PIN signal counts divided by square root of the total PIN counts (including the NXB) therein, and provides rough measure of the hard X-ray data quality.  Figure \ref{fig:HNR} plots the hardness against the PIN significance for the seven data sets.
  The hardness ratio is considerably higher in the other observations, of which at least the 2nd, 3rd, and the 4th ones were acquired in the hard state according to RMD11.
  Among them, the 3rd observation has the highest PIN significance with hardness comparable to those of the 2nd and 4th.  Therefore, we have selected for our detailed analyses the 1st and 3rd observations, as representative cases of the soft and hard states, respectively.  The other data sets will be analyzed in a future work, including particularly the 6th and 7th ones which are inferred to exhibit very hard spectra.

\subsection{Data Processing}\label{ss:DataProcessing}
\subsubsection{XIS data processing}

On September 28 when the source was in the soft state, the XIS was operated with 1/4 window and 0.5 s burst options.  We selected those XIS events of which the GRADE was 0, 2, 3, 4, or 6.  Time intervals after exits from the South Atlantic Anomaly were set to be longer than 436 s.
  Since the XIS count rate was so high as to need pile-up corrections, we discarded the image center ($< 1'$)  when accumulating on-source events as described above.  Background events were obtained from a larger annulus, which spans a radius range of $4'$ to $6'$.
A pile-up corrected and background-subtracted XIS FI (XIS0 + XIS3) count rate was 137.7 ct s$^{-1}$.
   The count rate in table \ref{tb:Obss} was obtained by restoring the counts to be observed within the excluded central region ($<1'$) with the help of the point-spread function \citep{Serlemitsos2007}; this process is expected to be accurate to $\sim 15$\% \citep{Yamada2011}.
  Due to the burst option, the net exposure of XIS0 and XIS3 was 3.5 ks each, which is a quarter of 13.8 ks shown in table \ref{tb:Obss}.

  In the other observations including the October 9 one, when the object was presumably in the hard state, the XIS was operated with the 1/4 window option, but without incorporating the burst option since the expected count rate was not very high.
  We accumulated the on-source events within a circle of $3'$ radius, because the pile-up correction was not required.  The background events were taken from an annulus with the inner and outer radii of $3'$ and $5'$, respectively.

  Light curves of the processed XIS data of the two observations (the 1st and 3rd), employed for our detailed study, are shown in figure \ref{fig:LCsoft}, panel (a) and (c).  In either observation, no burst-like events are seen and the time variation is less than $\sim$ 10\%.  Therefore, for our spectral analysis, we utilized time-averaged data, accumulated over a 0.8--10 keV range.
  To avoid the instrumental silicon K-edge and gold M-edge, where calibration uncertainties are large, we excluded two energy ranges, 1.7--1.9 keV and 2.2--2.4 keV, respectively.

\subsubsection{HXD data processing}\label{ss:222}
In all observations, the HXD data were screened by conditions that the time intervals after exits from the South Atlantic Anomaly should be longer than 500 s, the elevation of the target above the Earth limb should be $> 5^\circ$, and the geomagnetic cutoff rigidity should be higher than 8 GV.
  The Non X-ray Background (NXB) spectra were created from a fake-event file provided by the HXD team, and then subtracted from the on-source spectra.  In any observation, the Cosmic X-ray Background (CXB) was not subtracted from the data at this stage but included as a fixed model in the following spectral fits (section \ref{s:SA}).  The HXD-PIN count rates in table \ref{tb:Obss} were calculated before removing the CXB contribution.

\begin{figure*}[htbp]
  \begin{center}
    \FigureFile(80mm,80mm){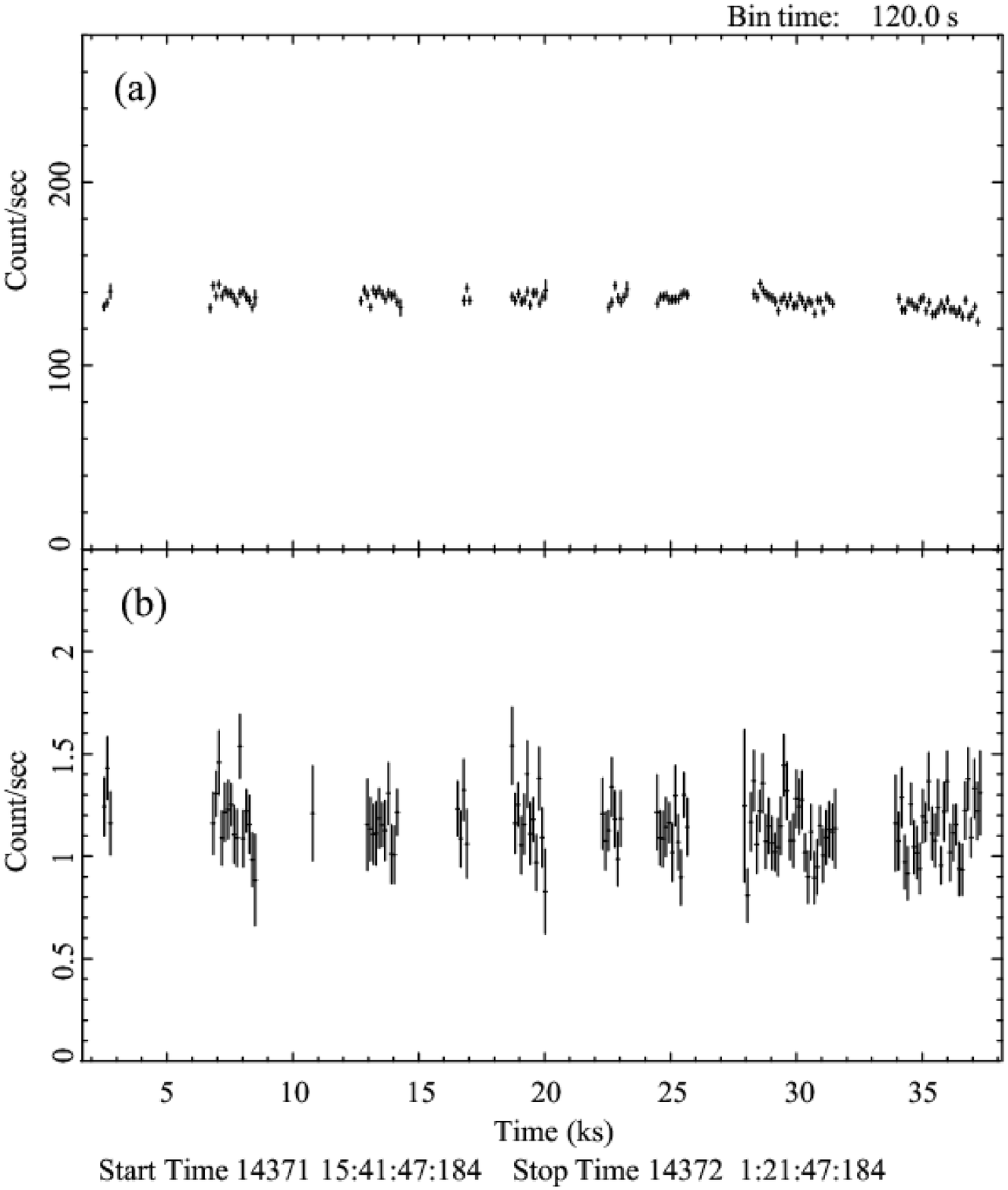}
        \FigureFile(80mm,80mm){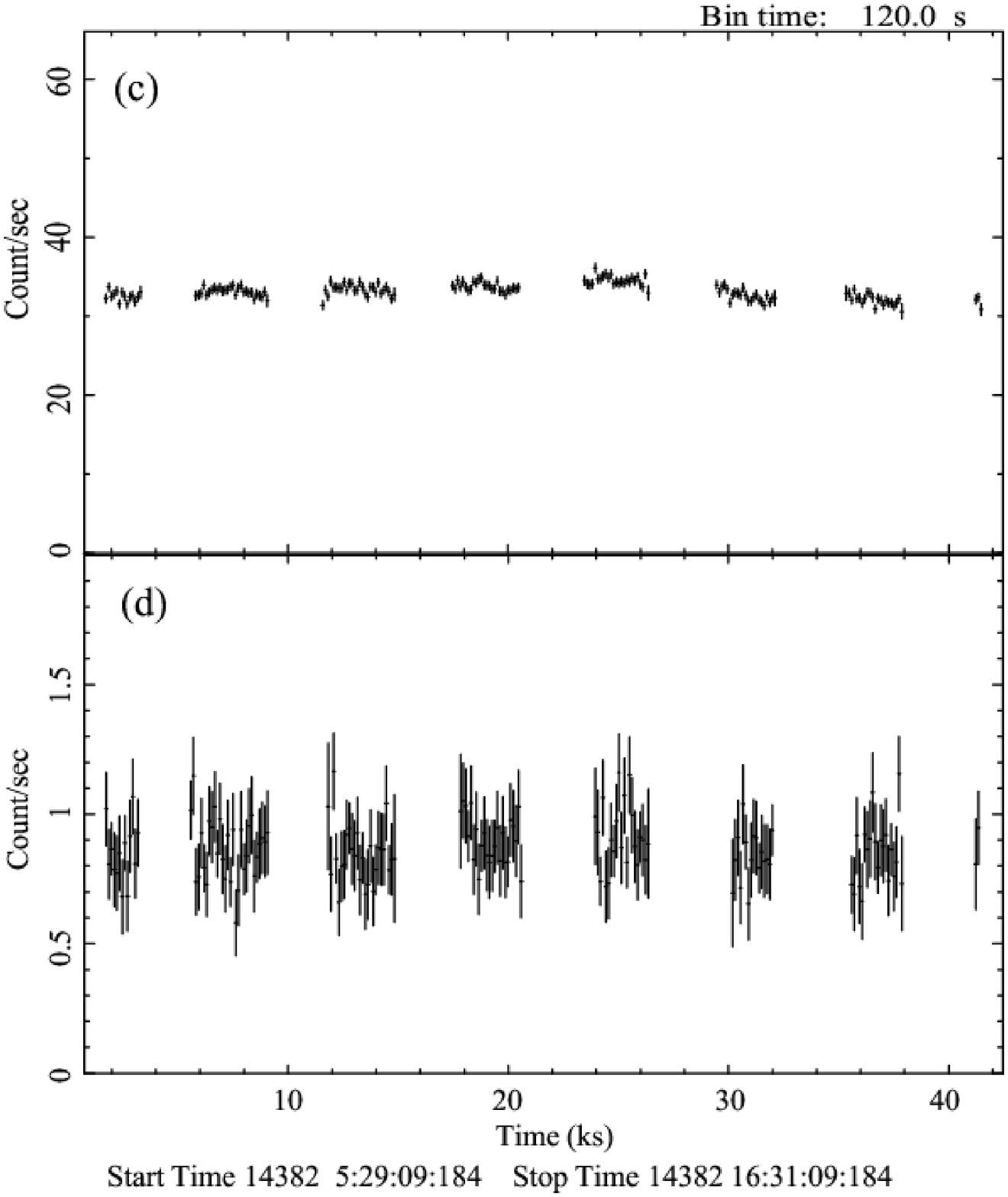}
  \end{center}
\caption{Light curves of Aquila X-1 with a 120 s binning obtained on 2007 September 28, in 0.8--10 keV with the XIS (XIS0 plus XIS3: panel a) and 12--50 keV with HXD-PIN (panel b). Panels (c) and (d) are the same as (a) and (b), respectively, but obtained on 2007 October 9.  The HXD-PIN light curves include the CXB, but not the NXB.  The rate in panel (a) is $\sim 50$\% of that in table \ref{tb:Obss}, due to the exclusion of the events within $1'$.
}
\label{fig:LCsoft}
\end{figure*}

  Analyzing the soft-state (September 28) data, 
 we detected positive source signals (after subtracting the NXB and CXB) in the energy band of 12--50 keV, but we quote the source detection conservatively in 12--31 keV, where the signal exceeded 5.6\% of the NXB which is a typical 2.5$\sigma$ error for an exposure of $\sim 14$ ks \citep[calculated from][]{Fukazawa2009}.  Further examination on this point is given in subsection \ref{sss:314}.
  The HXD-GSO data were consistent, within statistical errors, with the background over the whole energy range (50--600 keV).
  In the October 9 observation, which we have selected as the representative of the hard state, the HXD-PIN and HXD-GSO signals were successfully detected over 12--50 keV and 50--100 keV, respectively.
  The 50--100 keV HXD-GSO count rate was $0.19\pm 0.03$ ct s$^{-1}$.  
  This is significant, even considering systematic NXB error which is $\sim 0.6$\% (1$\sigma$) for the HXD-GSO case, or $\sim 0.045$ ct s$^{-1}$ for an exposure of $\sim 10$ ks \citep{Fukazawa2009}.
  HXD-PIN light curves of the selected two observations are shown in figure \ref{fig:LCsoft} panel (b) and (d).

\section{Spectral analysis} \label{s:SA}
\subsection{The soft state}\label{ss:SoftState} 
In performing quantitative model fitting to the September 28 spectra using XSPEC (version 12.6.0), we compensated for the reduced data accumulation region (discarding $< 1'$) by properly calculating an XIS arf that excludes the central circle within $1'$.
  To fit the HXD-PIN spectrum, the CXB was included in the model as a fixed component, expressed as a function of energy $E$ as
\begin{equation}
\textmd{CXB}(E) = 9.41\times 10^{-3}  \left(\frac{E}{1\textmd{keV}}\right)^{\hspace{-0.3em}-1.29}\hspace{-1.4em}\exp\left(-\frac{E}{40\textmd{keV}}\right)  \label{eq:cxb}
\end{equation}
where the unit is photons cm$^{-2}$s$^{-1}$keV$^{-1}$FOV$^{-1}$.  Although the CXB surface brightness in the HXD-PIN range has an uncertainty of about $\pm 15\%$ \citep[e.g.][]{Turler2010},
this is by an order of magnitude smaller than
 typical statistical errors in
our HXD-PIN spectra, and is hence negligible.  The cross-normalization difference between the XIS and the HXD can be considered by multiplying the model for HXD-PIN with a constant factor of 1.158, which was established accurately through calibrations using the Crab Nebula data
 obtained on many different occasions \citep{Ishida2006,Ishida2007,Maeda2008}.
   Its uncertainty, $\pm0.013$, is smaller than the statistical errors associated with the present HXD-PIN data, but later we allow it to vary.

\subsubsection{Fit with diskbb + blackbody}\label{ss:FitWithdBBbb}
\begin{figure*}[htbp]
  \begin{center}
     \FigureFile(80mm,80mm){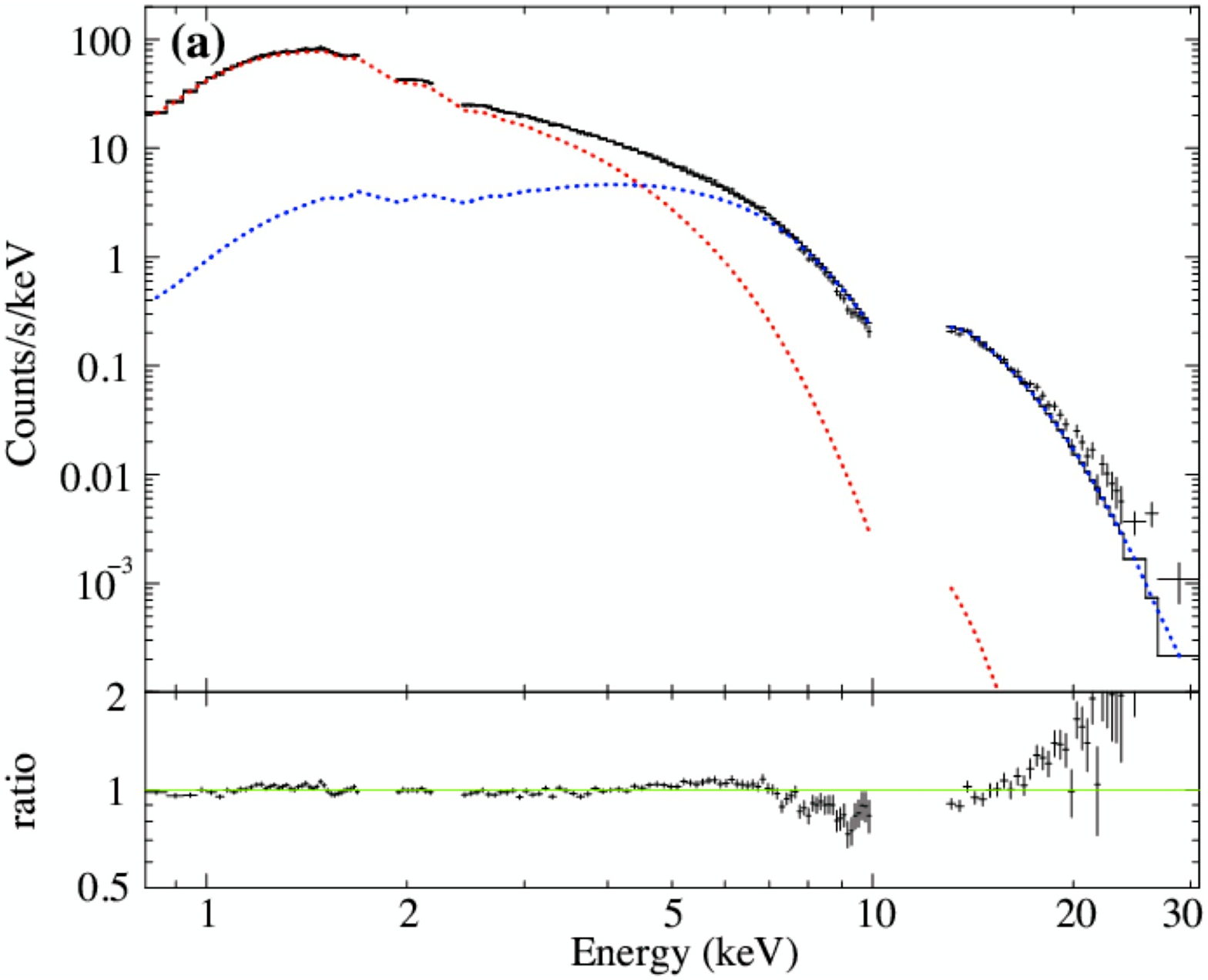}
     \FigureFile(80mm,80mm){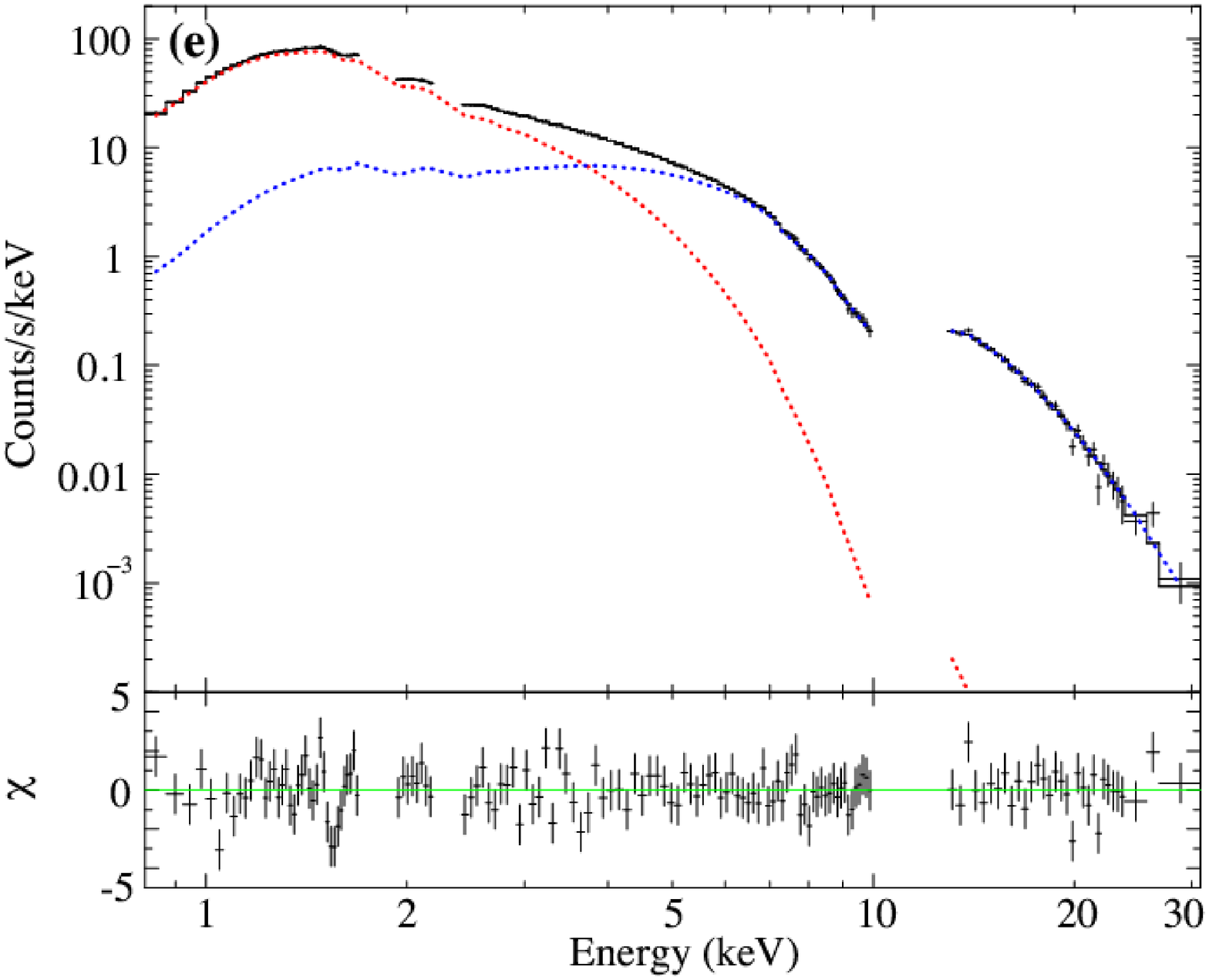}
  \end{center}
    \begin{center}
     \FigureFile(80mm,80mm){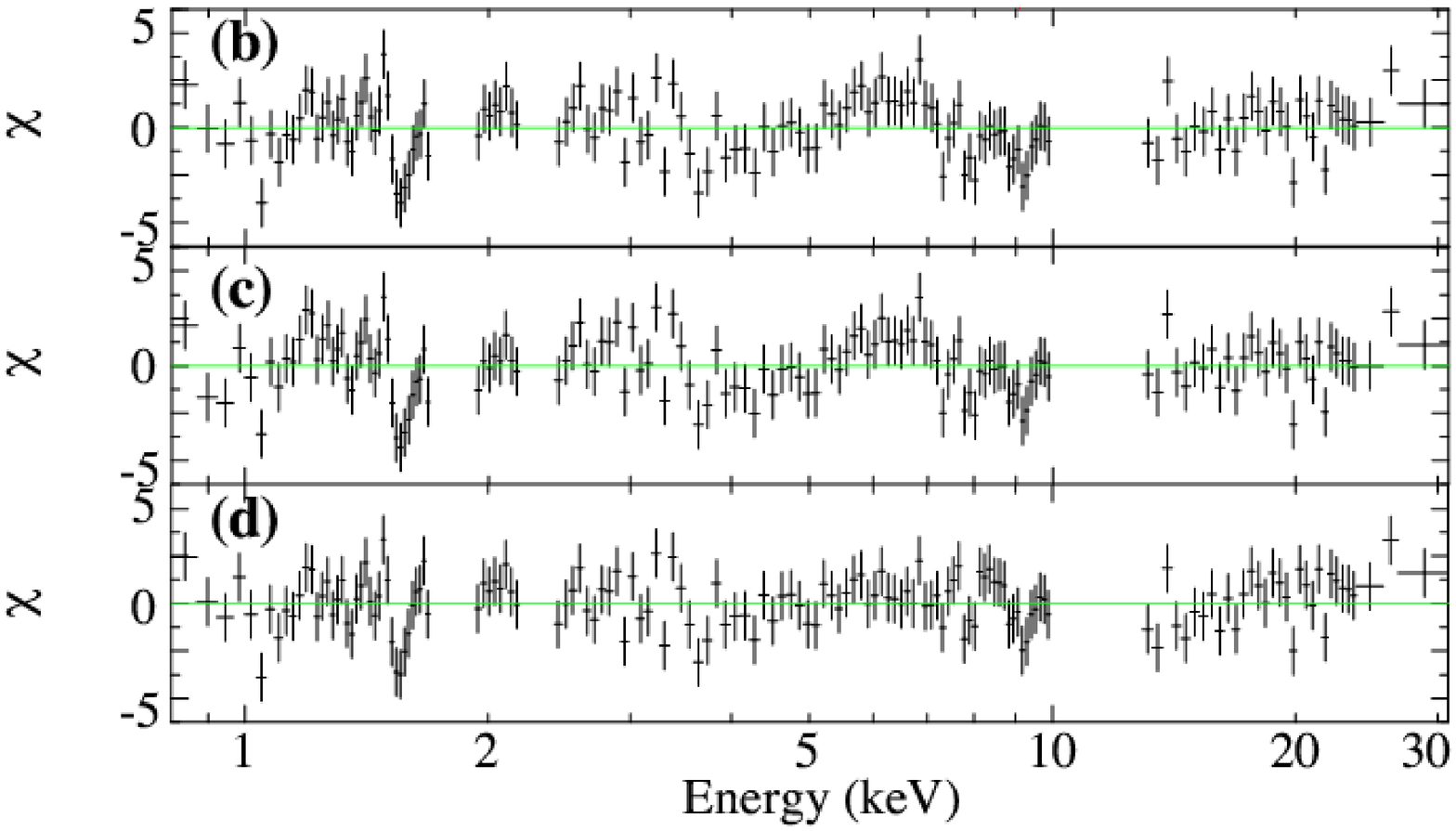}
     \FigureFile(80mm,80mm){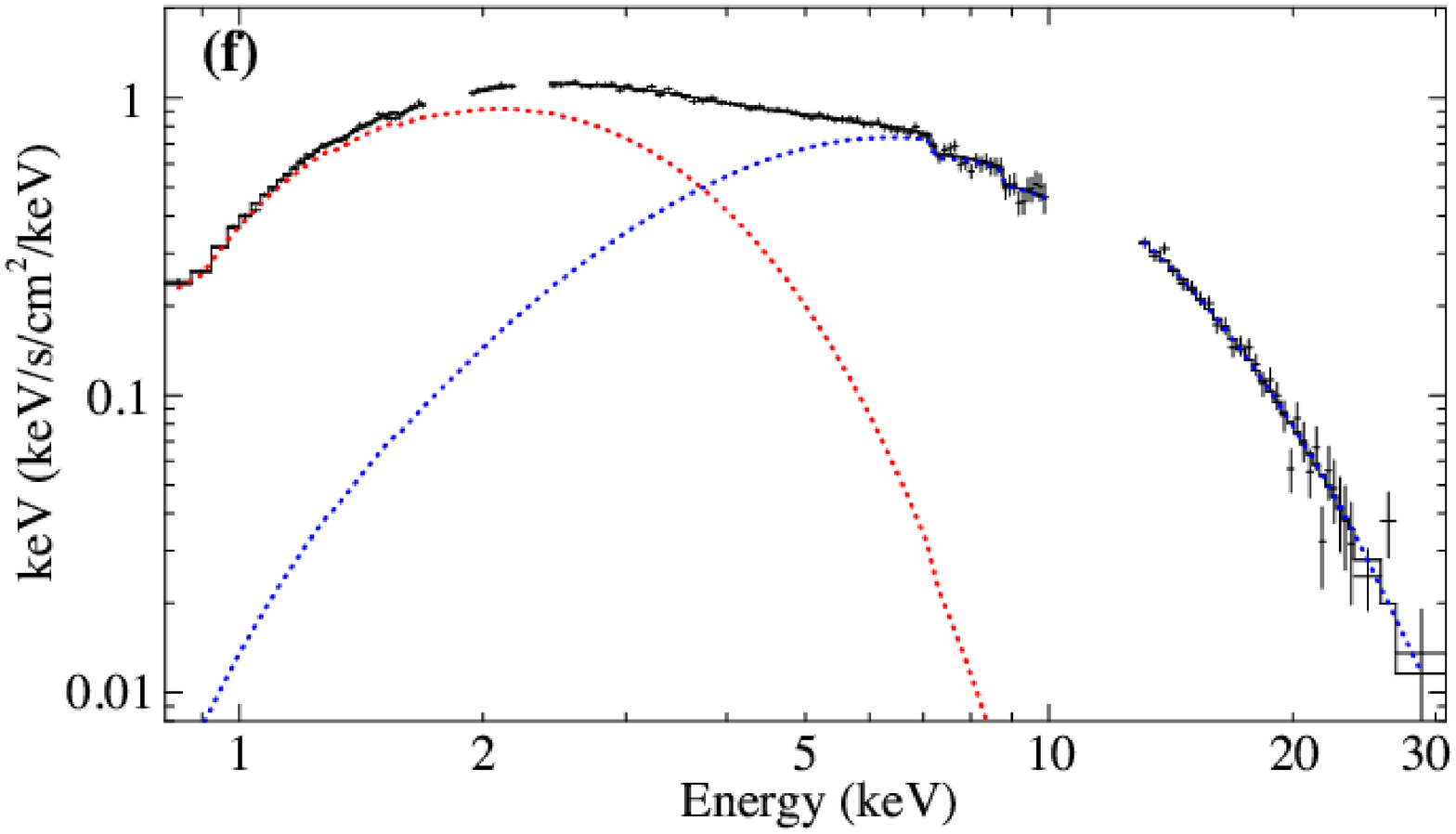}
  \end{center}
  \caption{(a) Simultaneous fitting of the XIS and HXD-PIN spectra in the soft state and its data-to-model ratio, using a {\tt diskBB} (red)  and a {\tt bbody} (blue) model.  (b) Residuals of the fit using a single {\tt diskBB} plus its Comptonization.  (c) The same as (b) but the  {\tt diskBB} source is replaced with a blackbody.  (d) The residuals incorporating three absorption edges to (b).
  (e) A fit with {\tt diskBB} plus {\tt nthcomp} ({\tt bbody}), and its residuals.  (f) The $\nu F \nu$ form corresponding to panel (e). The error bars represent statistical $1\sigma$ level.}
  \label{fig:bbdBBandcomp}
\end{figure*}

Figure \ref{fig:bbdBBandcomp} shows the XIS (FI) and HXD-PIN spectra obtained on September 28, presented after subtracting the NXB, but without removing the instrumental responses.
They indeed bear typical features of the soft state.
We jointly fitted them with a canonical model \citep{Mitsuda1984,Mitsuda1989,Makishima1989}, 
which consists of a soft disk blackbody component and a hard blackbody component.
  Hereafter, the CXB model as described above is included as a fixed component for the HXD-PIN spectra.
  The spectra were approximately reproduced as shown in figure \ref{fig:bbdBBandcomp}(a) and (b), in terms of the blackbody temperature of  $T_\textmd{\scriptsize{bb}} \sim 1.9$ keV, its radius of $R_\textmd{\scriptsize{bb}}\sim 2$ km, the inner disk radius of  $R_\textmd{\scriptsize{in}} \sim 11$ km, and the temperature therein  $T_\textmd{\scriptsize{in}} \sim 0.8$ keV.  However, the fit was not statistically acceptable with $\chi_\nu^2 = 3.24 (146)$.  Since $\chi_\nu^2$ is larger than 2.0, errors cannot be assigned to the parameters, and the residuals are presented in the form of data vs model ratio rather than in units of $\chi$.
  This model failure results from noticeable negative residuals around 7.0--9.0 keV and positive ones above 20 keV.

\subsubsection{Fit with an optically-thick component plus its Comptonization}

The positive fit residuals seen in figure \ref{fig:bbdBBandcomp}(a) above $\sim 15$ keV 
are suggestive of inverse-Compton effects, working possibly on blackbody as reported in previous studies \citep{Mitsuda1989,Gierlinski2002,Lin2007,Cocchi2011,Tarana2011,Takahashi2011}.
We hence tried, as a first-cut attempt, models involving a single optically-thick emission,
assuming that its fraction is Comptonized while the rest is directly visible.
As the thermal Comptonization model,
we employed {\tt nthcomp} \citep{Zdziarski1996,Zycki1999} here,
because it can select between a blackbody and a disk blackbody as the seed photon source,
and can handle a relatively wide range of optical depths.
  Thus, two models were constructed, 
{\tt diskBB+nthcomp(diskBB)}  and {\tt bbody+nthcomp(bbody)},
where the seed photon source to {\tt nthcomp} is indicated in parentheses.
The photon temperature was left free but constrained to be the same 
between the direct emission and the Comptonized one, 
while their normalizations were both left free,
as well as the optical depth of {\tt nthcomp}.
As a result, 
the former model gave $T_{\rm in}=0.59_{-0.03}^{+0.02}$ keV  with the direct fraction of $0.6\pm 0.1$ (60\% being direct), while the latter model  $T_{\rm bb}=0.38_{-0.2}^{+0.1}$ keV with the direct fraction of $0.3\pm 0.1$.
However, as presented in figure \ref{fig:bbdBBandcomp}(b) and figure \ref{fig:bbdBBandcomp}(c), neither fits were successful yet since they gave
$\chi^2_\nu (\nu) =1.56 (146)$ and $\chi^2_\nu (\nu) =1.60 (146)$, respectively.
These fits were not improved at all 
even when the cross-normalization between the XIS and HXD-PIN was allowed to vary;
$\chi^2_\nu (\nu) =1.57 (145)$ and $\chi^2_\nu (\nu) =1.61 (145)$.

The above fit failures are due, at least partially, to the noticeable negative residuals
seen in figure \ref{fig:bbdBBandcomp} (panels b and c) at energy ranges of 7.0--7.5 keV, 8.5--9.0 keV, and possibly 1.5--1.6 keV.
  To account for the two higher-energy features,
we hence added a neutral Fe K-edge and an ionized Fe K-edge, 
with their energies both left free.
  To eliminate that around 1.5--1.6 keV, 
which could arise from calibration errors of the  Al K-edge 
in the X-Ray Telescope (Serlemitsos et al. 2007) of Suzaku, 
another edge absorption was added with its energy fixed to 1.56 keV.
  Then, the {\tt diskBB+nthcomp(diskBB)} and {\tt bbody+nthcomp(bbody)} fits were 
significantly improved to $\chi^2_\nu (\nu) = 1.34 (141)$ and $1.27 (141)$, respectively.
Fit residuals of the former case are presented in figure \ref{fig:bbdBBandcomp}(d).
Taking the latter case  for example,
the two Fe-K  edge energies were obtained at
$7.2 \pm 0.1$ and $8.8 \pm 0.2$ keV,
corresponding to those of neutral Fe and He-like or somewhat lower ionization species,
with the associated optical depths of $0.10 \pm 0.04$ and $0.14 \pm 0.06$, respectively.
  Thus, the edges were confirmed to be significant
and their parameters are generally reasonable,
although neither model has achieved adequate fit goodness yet.

\subsubsection {Fit with a disk blackbody and a Comptonized blackbody}

Since the models with single seed photon population  (panels b, c, and d of figure \ref{fig:bbdBBandcomp}) were found to be unfavorable, 
we proceed to double  seed photon population models, 
in which the direct emission and the Comptonized emission 
have different optically-thick radiation sources.
As the most natural extension from the classical model utilized  in subsection \ref{ss:FitWithdBBbb} and presented in figure \ref{fig:bbdBBandcomp}(a),
a model of the form {\tt diskBB+nthcomp(bbody)} was constructed,
in which the two optically-thick components were allowed to take different temperatures,
$T_{\rm in}$ and $T_{\rm bb}$.
This describes a configuration
where the disk emission is directly visible
while the entire blackbody emission from the NS surface is Comptonized.
When the three edges are not considered, 
the fit gave $\chi^2_\nu (\nu) =1.44(145)$.
Although this is not yet acceptable,
it is significantly better (by $\Delta \chi^2 = -18.4 $ to $-24.1$ for $\Delta \nu=-1$) 
than those obtained in the previous subsection
under the single-source modelings.

When the three edges are incorporated, 
the fit  has become fully acceptable 
to $\chi^2_\nu (\nu) =1.15 (140)$,
with a null-hypothesis probability of 0.1.
The obtained fit is shown in figure \ref{fig:bbdBBandcomp} (panels e and f),
and the model parameters are summarized in table \ref{tb:cTTdBB-para};
the results virtually remained the same even
when leaving free the XIS vs. HXD cross-normalization.
The {\tt nthcomp} model specifies the Comptonization via two parameters, namely, the coronal electron temperature $T_\textmd{\scriptsize{e}}$, and the spectral slope $\Gamma_\textmd{\scriptsize{c}}$.
  In table \ref{tb:cTTdBB-para}, we have converted this $\Gamma_\textmd{\scriptsize{c}}$ (obtained as $3_{-1}^{+3}$) 
 to the optical depth of $\tau\sim 6$, using relations as
\begin{eqnarray}
\Gamma &=& -\frac{1}{2} + \sqrt{\frac{9}{4}+\frac{1}{\frac{kT_\textmd{\scriptsize{e}}}{m_\textmd{\scriptsize{e}}c^2 }\tau(1+\frac{\tau}{3})}}
\end{eqnarray}
\citep{Lightman1987}.  For reference, the Compton $y$-parameter, $y\equiv(4kT_\textmd{\scriptsize{e}}/m_\textmd{\scriptsize{e}}c^2)\times \textmd{max}(\tau,\tau^2)$, becomes $0.8\pm0.1$.

  In table \ref{tb:cTTdBB-para} and hereafter, 
the  innermost disk radius $R_{\rm in}$ has been derived from the {\tt diskBB} model, 
assuming that its radius parameter is in fact equals to  $\xi \kappa^2 R_{\rm in} \sqrt{\cos i}$.
Here, $\xi=0.412$ is a correction factor for the inner boundary condition
\citep{Kubota1998,Makishima2000}, 
$\kappa =1.7$ is the standard color hardening factor \citep{Shimura_Takahara1995},
and  $i$ is the disk inclination which we assumed to be $45^\circ$ for simplicity.
Thus, the estimated emission-region radius can be
reasonably identified with a fraction of the NS surface.

  Just for reference,
we examined another possible model configuration,
namely {\tt bbody + nthcomp(diskBB)}.  Like in the case of table \ref{tb:cTTdBB-para}, all model parameters (again except edge1 energy at 1.56 keV) were left free.
  It yielded $\chi^2_\nu (\nu) =1.53 (145)$ and $1.30 (140)$,
without and with the edges, respectively.
The fit was not improved [$\chi^2_\nu (\nu) =1.29 (139)$] 
even when the cross-normalization was left free.
We no longer consider this alternative modeling,
since these fits are significantly worse 
than those obtained with {\tt diskBB+nthcomp(bbody)},
and are not acceptable.

\subsubsection {Significance of the hard tail}\label{sss:314}
\begin{figure*}[htbp]
  \begin{center}
     \FigureFile(80mm,80mm){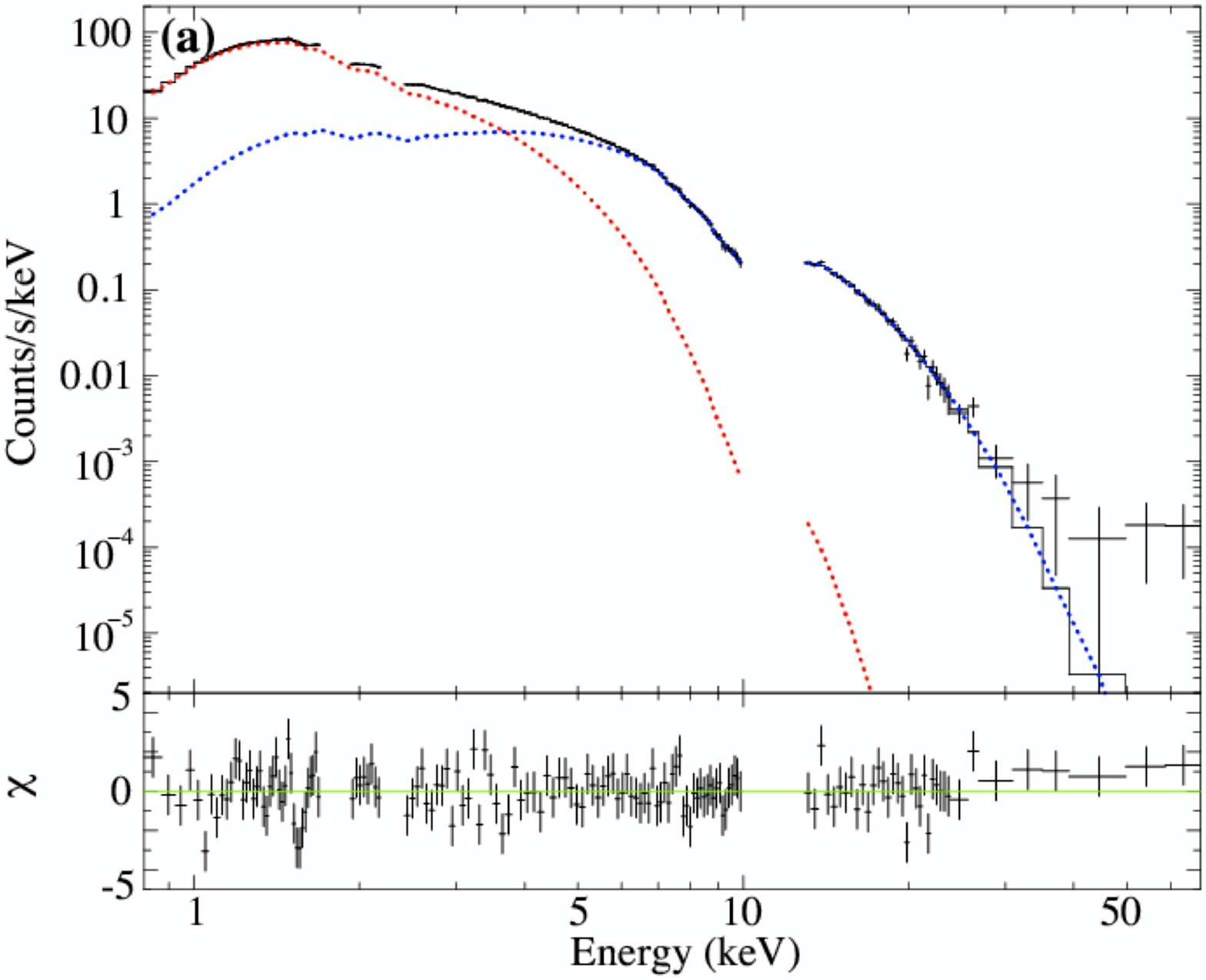}
     \FigureFile(80mm,80mm){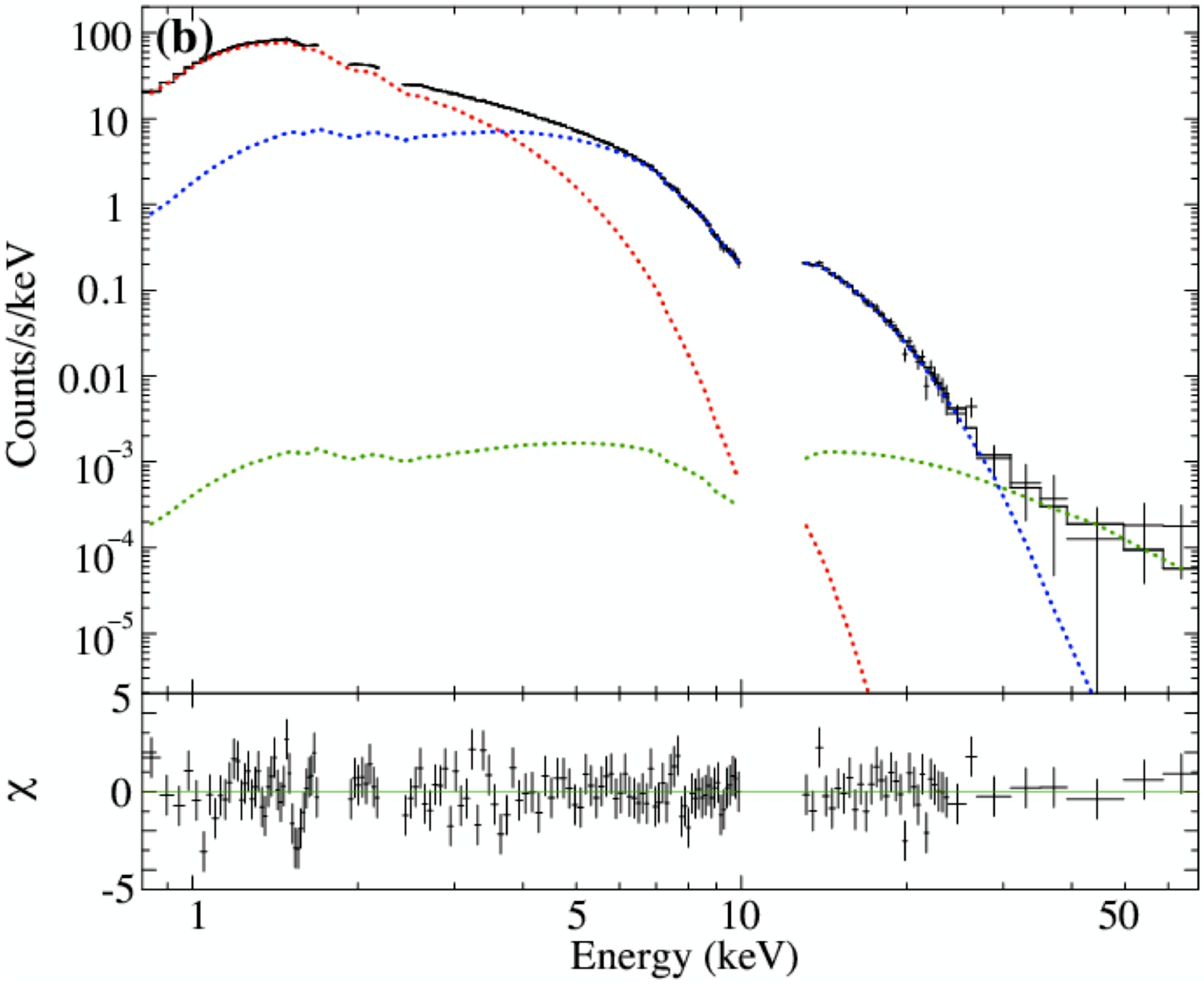}
     \FigureFile(80mm,80mm){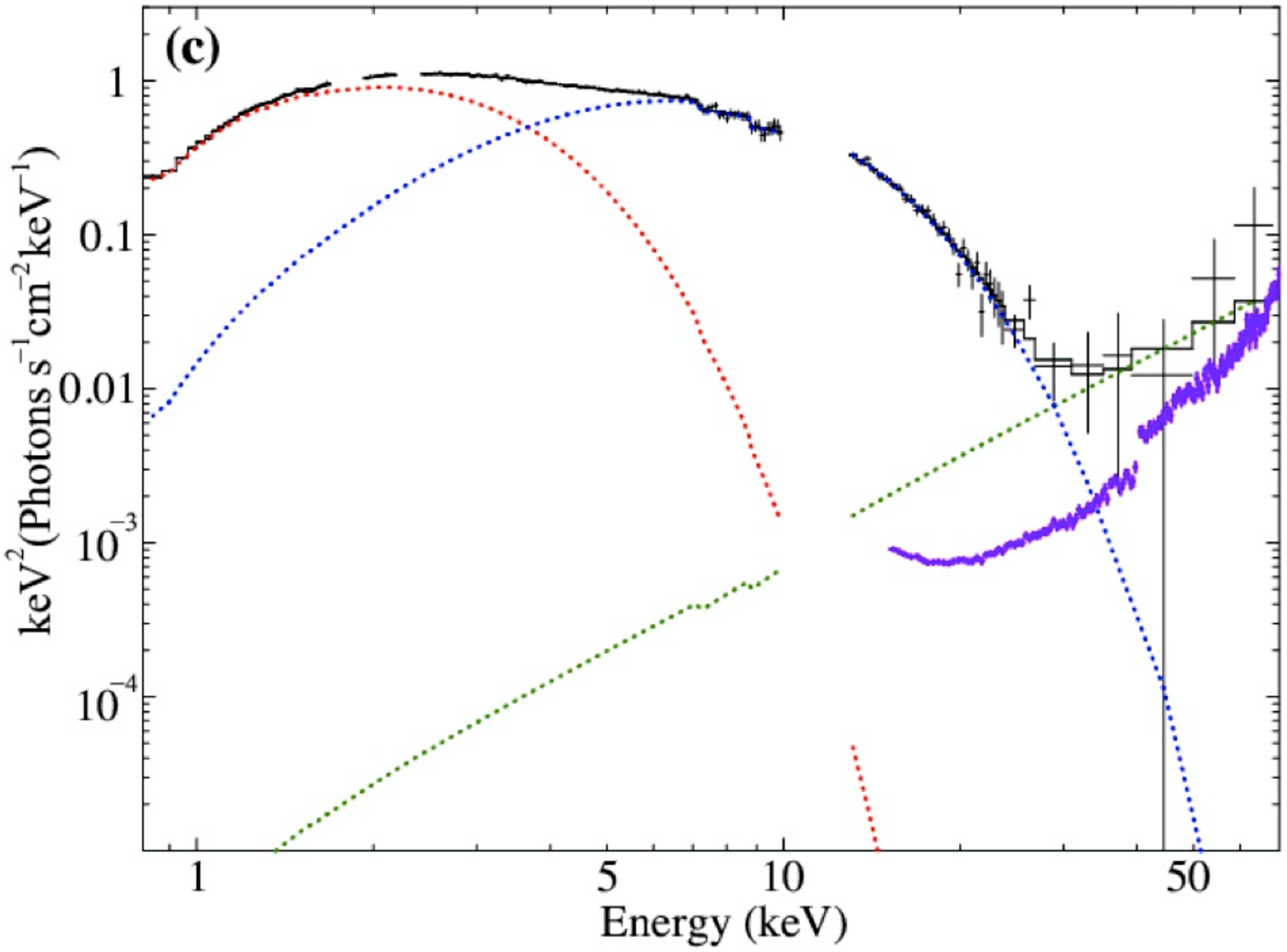}
  \end{center}
  \caption{The same soft-state spectra as figure \ref{fig:bbdBBandcomp}, but further including five additional HXD-PIN data points in the 31--70 keV range.  (a) The {\tt diskBB}+{\tt nthcomp}({\tt bbody}) fit, which is nearly identical to that in figure \ref{fig:bbdBBandcomp} (e), but the parameters were re-optimized.
    (b) The same as panel (a), but adding a PL (green dotted line) with a fixed slope of $\Gamma = 0.0$. 
    (c) The $\nu F \nu$ form of panel (b), where systematic $1\sigma$ uncertainties of the NXB (see text) are superposed in purple.
     Again, the error bars of the data points represent statistical $1\sigma$ range.
  }
  \label{fig:HTs}
\end{figure*}

 Analyzing the soft-state  Suzaku data on September 28, 
RMD11 reported the detection of a hard-tail component
with a very flat slope of $\Gamma \sim 0$,
which appeared in the 30--70 keV range of the HXD-PIN spectrum.
To examine its reality, we extended the upper energy bound of the HXD-PIN data
from the previously employed 31 keV to 70 keV,  by adding five data points.
As shown in figure \ref{fig:HTs} (a),
these additional data points indeed exceed the best-fit 
{\tt diskBB}+{\tt nthcomp}({\tt bbody}) model (with the three edges),
even when the model is refitted.
Then, by adding a PL with $\Gamma=0$ (fixed),
the excess was explained away as seen in figure \ref{fig:HTs} (b),
and the fit goodness changed slightly from $\chi^2_\nu(\nu)=1.15 (145)$ to $1.12 (144)$
 with $\Delta \chi^2 = -5.9$ for $\Delta \nu= -1$.
Within 90\% errors, the obtained PL normalization, $(9\pm 6) \times 10^{-6}$,
agrees with $(2.1\pm 1.5) \times 10^{-5}$ reported by RMD11.
According to an $F$-test, 
the probability for this fit improvement to arise by chance is about 2\%,
even though the statistical deviation of the additional five data points 
from the best-fit {\tt diskBB}+{\tt nthcomp}({\tt bbody}) model
yields a relatively low reduced chi-square of $\chi_\nu^2 = 1.25 ( \nu= 5)$. 
When allowed to vary,
$\Gamma$ was not well constrained,
and floated over a rather large range, e.g., 0--3.0.
In any case, the other fit parameters were virtually unaffected, 
and remained consistent with table \ref{tb:cTTdBB-para}.

The above estimates considered only statistical errors.
However, we must also consider systematic uncertainties in the HXD-PIN data points.
The NXB modeling in the 40--70 keV range of HXD-PIN is known 
to have $1 \sigma$ systematic errors of 2.8\% and 1.8\%,
for an exposure of 10 ks and 20 ks, respectively \citep{Fukazawa2009}.
Since the present data have an exposure of 14 ks, 
we may simply estimate the $1 \sigma$ NXB error as  2.2\%,
as the geometric mean of 2.8\% and 1.8\%.
Similarly, we assign a $1\sigma$ systematic NXB error of 1.2\% below 40 keV.
Figure \ref{fig:HTs}(c) shows the same spectrum as figure \ref{fig:HTs}(b), but in the $\nu F\nu$ form,
  in comparison with these $1\sigma$ NXB uncertainties. 
  When this systematic error is included in the fit goodness evaluation,
the $F$-test chance probability for the additional PL component increased to $\sim 7\%$.
Therefore, the hard tail is significant at the 90\% confidence level, but not at 95\%.

\begin{table}[htbp]
\caption{Parameters of the {\tt diskBB} plus {\tt nthcomp} fit for the soft state.\footnotemark[$*$] }
\centering
\begin{minipage}{8cm}
\begin{tabular}{ccc}\hline
Component & Paramater & Value \\\hline
wabs & $N_\textmd{\scriptsize{H}}$ ($10^{22}$cm$^{-2}$) & $0.36 \pm 0.01$ \\\hline
edge1 & $E_\textmd{\scriptsize{edge}}$ (keV) & 1.56 (fixed) \\
	    & optical depth  					& 0.03 $\pm$ 0.01 \\\hline
edge2 & $E_\textmd{\scriptsize{edge}}$ (keV) & $7.1 \pm 0.1$ \\
	    & optical depth  					& $0.14 \pm 0.04$ \\\hline
edge3 & $E_\textmd{\scriptsize{edge}}$ (keV) & 8.7 $\pm$ 0.2 \\
	    & optical depth 					& 0.13 $\pm$ 0.06 \\\hline
diskBB & $T_\textmd{\scriptsize{in}}$ (keV) & $0.73_{-0.03}^{+0.02}$ \\
	    & $R_\textmd{\scriptsize{in}}$ (km)\footnotemark[$\dagger$]  & $15\pm 2$ \\\hline
nthcomp& $T_\textmd{\scriptsize{bb}}$ (keV) & $1.4_{-0.1}^{ +0.2}$ \\
	      & $T_\textmd{\scriptsize{e}}$ (keV) & $3_{-2}^{+28}$ \\
	      & optical depth 	& $6\pm 4$ \\
	      & $R_\textmd{\scriptsize{bb}}$  (km) & $3.0_{-0.7}^{+0.8}$ \\\hline
Fit goodness & $\chi_\nu^2(\nu)$ & 1.15 (140)\\\hline
\end{tabular}
\label{tb:cTTdBB-para}
\footnotetext[$*$]{Errors represent 90\% confidence limits.}
\footnotetext[$\dagger$]{After applying the corrections described in the text, and assuming a distance of 5.2 kpc.}
\end{minipage}
\end{table}

\subsection{The hard state} 
The October 9 spectra, detected over the 0.8--100 keV band with the XIS, HXD-PIN, and HXD-GSO (subsection \ref{ss:222}), were analyzed in the same way.  As presented in figure \ref{fig:HNR} and reported by RMD11, the spectra on this occasion show typical characteristics of the hard state; we therefore consider that the source was actually in the hard state on this occasion.  Since the interstellar absorption, corresponding to {\tt wabs} in table \ref{tb:cTTdBB-para}, was obtained as $N_\textmd{\scriptsize{H}} = 0.36 \times 10^{22}$ cm$^{-2}$ in the soft state analyses where the soft band flux is higher, we fixed $N_\textmd{\scriptsize{H}}$ to this value in the hard state analyses.  Neither of the three edges were incorporated, because the soft band flux was relatively low in this observation.
  The model normalization was constrained to be the same between HXD-PIN and HXD-GSO \citep{Kokubun2007}.

\subsubsection{Fit with a single Comptonized component}\label{sss:FitWithCompPS}
Since hard-state spectra are generally considered to be more dominated by Comptonization than the soft-state ones analyzed in subsection \ref{ss:SoftState}, we begin with fitting the October 9 spectra with a single Comptonized emission model, represented by  {\tt compPS} \citep{Poutanen1996}.  
  The use of this model, instead of {\tt nthcomp} used in subsection \ref{ss:SoftState}, is because it can more properly handle relativistic effects (e.g. using the Klein-Nishina formula instead of the Thomson cross section), and hence is more suited to the hard-state spectrum which extends to $\sim 100$ keV.
  Actually, this model was successfully employed by \cite{Takahashi2008} and \cite{Makishima2008}, in the study of GRO 1655$-$40 and Cygnus X-1, respectively, both observed with Suzaku in the hard state.
  Hereafter, a relatively narrow iron line is added to the fitting model.  
As to  seed photons of the inverse Compton emission, we assumed either a blackbody with free $T_\textmd{\scriptsize{bb}}$ or a diskBB with free $T_\textmd{\scriptsize{in}}$.
  As shown in figure \ref{fig:bbdBBcPS}(a),
  the spectra were roughly reproduced by a strong Comptonization operating on either {\tt bbody} or {\tt diskBB} (though only the former is shown in the figure), 
with $T_\textmd{\scriptsize{bb}} \sim 0.5$ or $T_\textmd{\scriptsize{in}} \sim 0.6$ keV, respectively.
  However both of them were unacceptable, mainly due to positive residuals below $\sim$4 keV.

  When leaving $N_\textmd{\scriptsize{H}}$ free, the {\tt compPS}({\tt bbody}) fit was much improved from $\chi^2_\nu(\nu)=8.64 (186)$ to $1.85 (185)$, but the result was not yet acceptable because of the structures below $\sim 4$ keV.
By this additional degree of freedom, the {\tt compPS}({\tt diskBB}) fit improved from $\chi^2_\nu(\nu)=2.52 (186)$ to $1.16 (185)$.
  However, the derived value of $N_\textmd{\scriptsize{H}}=0.277_{-0.1}^{+0.05}\times 10^{22}$ cm$^{-2}$ is not consistent with that of the soft state (table \ref{tb:cTTdBB-para}),
  and the obtained value of $R_\textmd{\scriptsize{in}} = 4\pm 1$ km is too small.
Leaving free the XIS vs HXD cross normalization did not improve either fit.
  Therefore, the spectrum cannot be reproduced by a single thermal Comptonization model, regardless of the choice of the seed photon source.  More specifically, the data are suggestive of the presence of an additional component at the softest spectral end, and a reflection hump at $\sim 30$ keV.

\subsubsection{Fit with an optically-thick component plus its Comptonization}\label{sss:FitWithOptthick+CompPS}
To account for the soft excess left by the single {\tt compPS} fits in figure \ref{fig:bbdBBcPS}(a), we modified the model, like in the soft-state analysis, and considered a case where a fraction of the seed photon source (either {\tt bbody} or {\tt diskBB}) is directly visible.
  Hence the model is {\tt bbody}+{\tt compPS}({\tt bbody}) or {\tt diskBB}+{\tt compPS}({\tt diskBB}).
  Again, the direct and Comptonized photons were assumed to have the same seed-photon temperature.
  However, in both cases, the fraction of directly seen component became so small ($\sim$0.04 and $\sim$0.001  for {\tt diskBB} and {\tt bbody}, respectively),
that the results were nearly the same as in the previous subsection.
This is mainly because the soft excess, appearing in $\leq 1$ keV, is rather softer than the Compton seed photon source, of which the photon temperature is specified as $\sim 0.5$ keV by the $\sim 1$ to $\sim 4$ keV part of the spectrum. 
 In short, the spectra cannot be reproduced by a single photon source, even allowing its fraction to be directly visible.

\subsubsection{Fit with an optically-thick component plus a separate Comptonized emission}\label{ss:323}
After previous works \citep{Christian1997,Church2001,Gierlinski2002}, as well as the results of subsection \ref{sss:FitWithOptthick+CompPS}, we proceed to a fit in which the directly-visible soft thermal emission is different from the seed source for {\tt compPS}.

\begin{figure*}[htbp]
\begin{center}
	\FigureFile(80mm,80mm){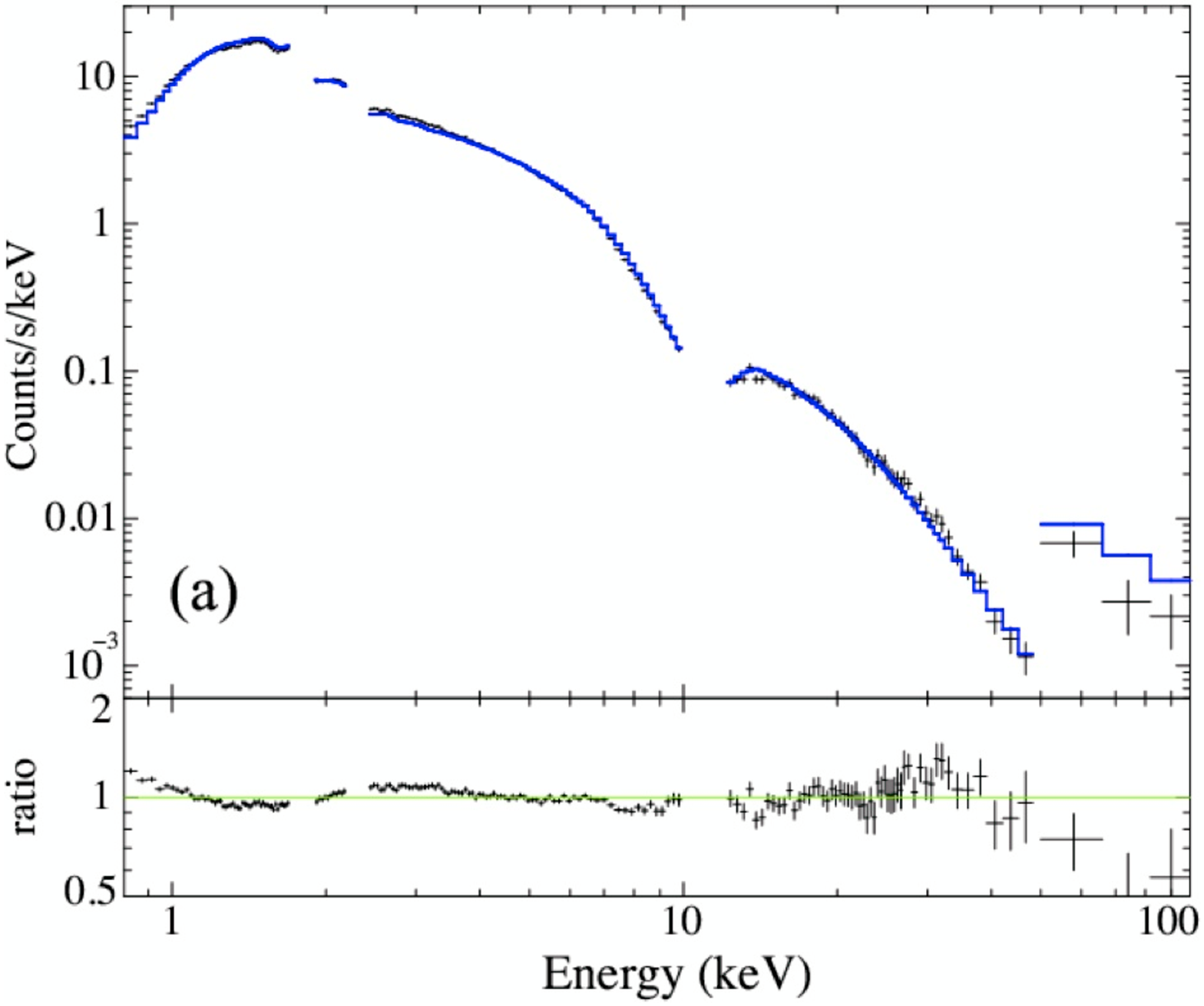}
	\FigureFile(80mm,80mm){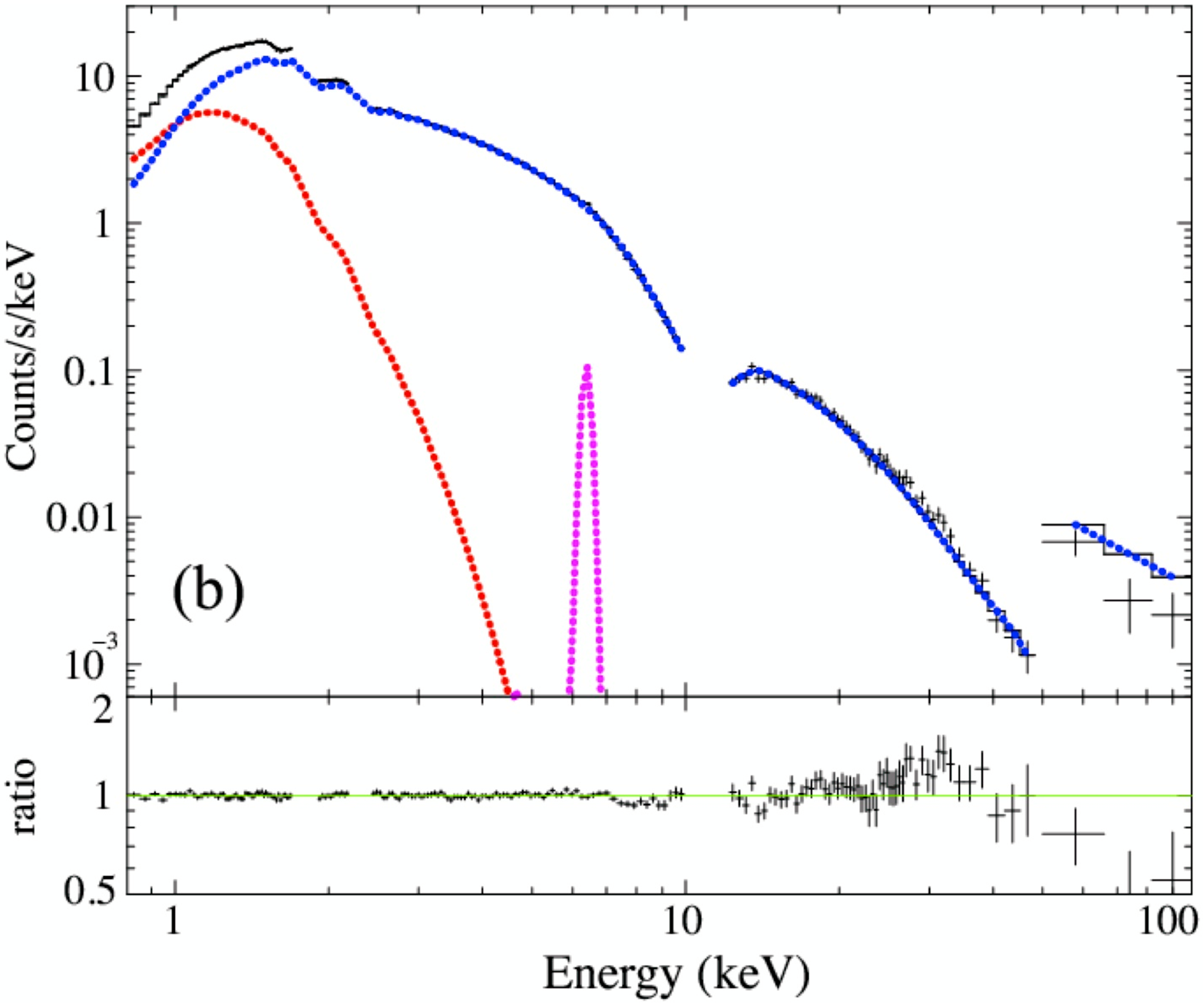}
  \end{center}
\begin{center}
   	 \FigureFile(80mm,80mm){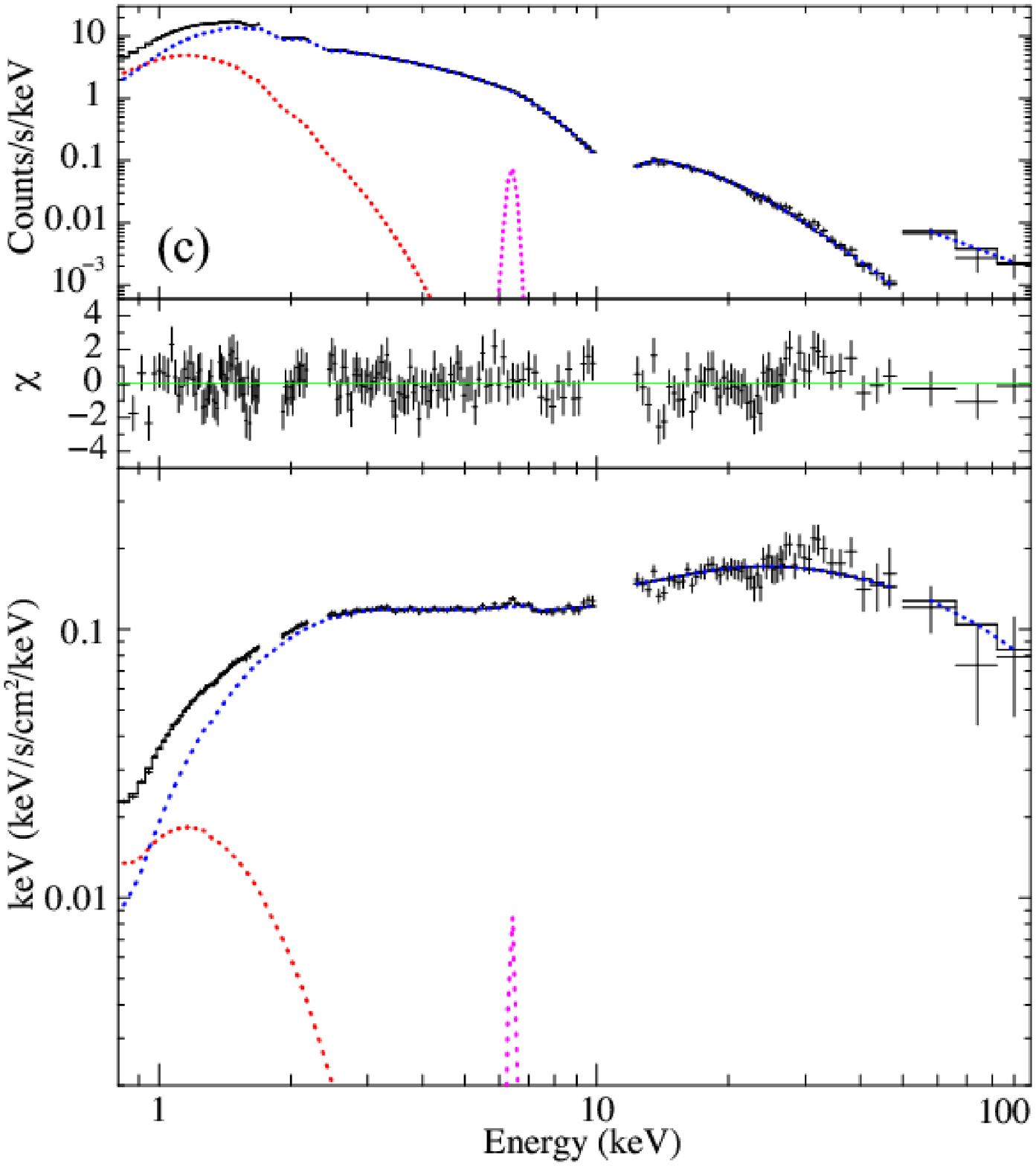}
   	 \FigureFile(80mm,80mm){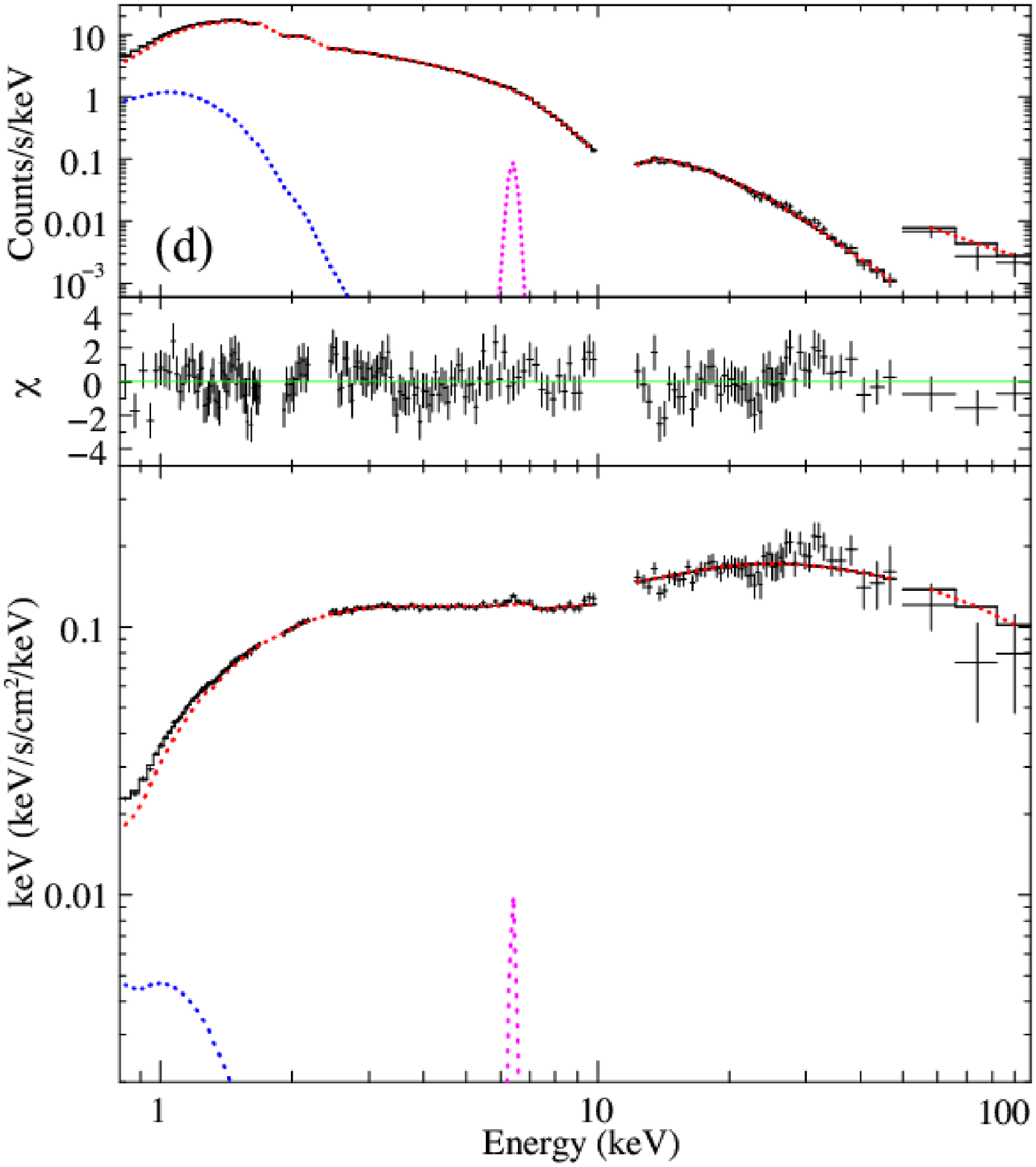}
  \end{center}
  \caption{Simultaneous fitting of the XIS, HXD-PIN, and HXD-GSO spectra in the hard state.  (a) A fit with {\tt compPS} ({\tt bbody}) and its data-to-model ratio.  (b) A fit with {\tt diskBB}+{\tt compPS} ({\tt bbody}) and its ratio.  (c) A fit with the same model as (b), but incorporating reflection.  The middle panel shows residuals, and the bottom one $\nu F\nu$ form of the fit.
  (d) A fit obtained by exchanging the seed photon and the optically-thick emission in (c), namely, with the {\tt bbody} + {\tt compPS} ({\tt diskBB}) model. 
  }\label{fig:bbdBBcPS}
\end{figure*}

  As one of such model forms, we tried {\tt diskBB}+{\tt compPS}({\tt bbody}) model, which has essentially the same form as our final model for the soft state (except Fe-K edges and lines).  
  As shown in figure \ref{fig:bbdBBcPS}(b), the soft-band data below $\sim 2$ keV have been well reproduced (hence with little merit of leaving $N_\textmd{\scriptsize{H}}$ free), but the fit was not yet acceptable with $\chi^2_\nu(\nu)=1.43 (184)$, due to a step-like feature in 7--10 keV and a hump at $\sim 30$ keV.
  These features were not reduced significantly by changing the XIS vs HXD cross normalization.
  Since they are suggestive of the presence of a reflection component like in the soft-state spectra, we incorporated a reflection embedded in {\tt compPS}.
  Then, as shown in figure \ref{fig:bbdBBcPS}(c) and table \ref{tb:bbdBBcPS-para},
the fit has become acceptable even when $N_\textmd{\scriptsize{H}}$ is fixed.  The obtained parameters, including the reflection solid angle, are reasonable for the modeling.

  As an alternative attempt, we fitted the spectra with the other model form, namely, {\tt bbody}+{\tt compPS}({\tt diskBB}).  Like the previous model, this fit was unacceptable without considering reflection,  $\chi^2_\nu(\nu)=2.05 (184)$, even when leaving $N_\textmd{\scriptsize{H}}$ and the cross normalization free.  When reflection is incorporated, the fit has become acceptable (with $N_\textmd{\scriptsize{H}}$ fixed), as shown in figure \ref{fig:bbdBBcPS} (d) and table \ref{tb:bbdBBcPS-para}.  
  The final fit goodness is somewhat worse than that of the  {\tt diskBB}+{\tt compPS} ({\tt bbody}) modeling with the fixed $N_\textmd{\scriptsize{H}}$.
  Further comparison between the successful two models is presented in section \ref{ss:43}.


\begin{table*}[htbp]
\caption{Fit parameters of  an optically-thick component plus a Comptonized emission for the hard state.}
\centering
\begin{minipage}{14cm}
\begin{tabular}{cccccc}\hline
Component & & \multicolumn{2}{c}{{\tt diskBB}+{\tt compPS}({\tt bbody})} & \multicolumn{2}{c}{{\tt bbody}+{\tt compPS}({\tt diskBB})} \\\hline
$N_\textmd{\scriptsize{H}}$ ($10^{22}$cm$^{-2}$) & & 0.36 (fixed) & $0.45_{-0.07}^{+0.05}$ & 0.36 (fixed) & $0.50\pm 0.06$ \\\hline
Opt. thick & model & diskBB & diskBB & BB & BB\\
		& $T_\textmd{\scriptsize{bb}}/T_\textmd{\scriptsize{in}}$ (keV) & $0.278\pm 0.02$ & $0.23_{-0.02}^{+0.03}$ & $0.16\pm 0.01$ & $0.144_{-0.007}^{+0.009}$ \\
		& $R_\textmd{\scriptsize{bb}}/R_\textmd{\scriptsize{in}}$ (km) & $21\pm 4$ & $41_{-14}^{+18}$ & $41\pm 8$ & $100\pm 30$ \\\hline
compPS	& seed & BB & BB & diskBB & diskBB\\
		& $T_\textmd{\scriptsize{in}}/T_\textmd{\scriptsize{bb}}$ (keV) & $0.51\pm 0.02$ & $0.48_{-0.02}^{+0.03}$ & $0.749_{-0.004}^{+0.003}$ & $0.68\pm 0.04$ \\
		& $R_\textmd{\scriptsize{in}}/R_\textmd{\scriptsize{bb}}$ (km) & $10\pm 2$ & $12\pm 3$ & $4.2\pm 0.6$ & $4.3\pm 0.8$ \\
		& $T_\textmd{\scriptsize{e}}$ (keV) & $35_{-5}^{+4}$ & $37_{-4}^{+5}$ & $52.4_{-0.9}^{+0.7}$ & $55_{-8}^{+7}$ \\
		& $\tau$ & $\geq 2.5$\footnotemark[$\dagger$]  & $2.5_{-0.2}^{+0.3}$ & $1.78_{-0.02}^{+0.01}$ & $1.6_{-0.2}^{+0.3}$ \\
		& refl\footnotemark[$*$] ($\Omega/2\pi$) & $0.6_{-0.1}^{+0.2}$ & $0.7_{-0.2}^{+0.2}$ & $0.60\pm 0.05$ & $0.6_{-0.1}^{+0.2}$ \\\hline
gaussian 	 & $E$ (keV)& 6.4 (fixed) & 6.4 (fixed) & 6.4 (fixed) & 6.4 (fixed) \\
		 & $\sigma$ (keV) & 0.1 (fixed) & 0.1 (fixed) & 0.1 (fixed) & 0.1 (fixed)  \\
		 & EW\footnotemark[$\ddagger$] (eV) & $22_{-16}^{+13}$ & $21_{-16}^{+14}$ & $23_{-16}^{+15}$ & $22\pm 17$\\\hline
fit goodness & $\chi_\nu^2 (\nu)$ & $1.02\: (183)$ & $1.00\: (182)$ & $1.10\: (183)$ & $1.04\: (182)$ \\\hline
\end{tabular}
\footnotetext[$*$]{Reflection solid angle.}
\footnotetext[$\dagger$]{The upper limit is larger than 3, which is the model limit in {\tt compPS}.}
\footnotetext[$\ddagger$]{Equivalent width.}
\end{minipage}

\label{tb:bbdBBcPS-para}
\end{table*}

\section{Discussion} 

\subsection{The two spectral states} 
We analyzed two out of the 7 Suzaku data sets of Aquila X-1, acquired in an outburst over 2007 September to October.  On September 28, when the source was in the typical soft state as expected from figure \ref{fig:HNR}, the signals were successfully detected over 0.8--31 keV with the XIS and HXD-PIN.  On October 9, in the typical hard state (figure \ref{fig:HNR}), 
the spectrum became much harder, and the signal detection was achieved over a broader range of 0.8--100 keV, with the XIS, HXD-PIN, and HXD-GSO.
  The soft-state luminosity in 2--20 keV and that of the hard state in 2--100 keV were $1.5\times 10^{37}$ erg s$^{-1}$ and $2.5\times 10^{36}$ erg s$^{-1}$, respectively, assuming a distance of 5.2 kpc and isotropic emission.  These values correspond to $\sim$8\% and $\sim$1\% of the Eddington luminosity, respectively, for a neutron-star mass of 1.4 $M_\odot$.  Therefore, the present results are consistent with the understanding that NS-LMXBs, in a broadly similar way to BHBs, make transitions between the soft and hard states at a threshold luminosity of several percent of the Eddington luminosity \citep{Bloser2000,Maccarone2003}.  Below, we discuss the two states separately, referring to the results obtained with $N_\textmd{\scriptsize{H}} = 0.36 \times 10^{22}$ cm$^{-2}$ and the cross normalization of 1.158 both fixed.

\subsection{Geometry in the Soft state} 
Using the parameters derived from the successful fit with {\tt diskBB}+{\tt nthcomp}({\tt bbody}) shown in figure \ref{fig:bbdBBandcomp}(e), we obtained a blackbody radius of $R_\textmd{\scriptsize{bb}} = 3.0_{-0.8}^{+0.7}$ km and an inner disk radius of $R_\textmd{\scriptsize{in}} = 15\pm 2$ km (table \ref{tb:cTTdBB-para}).
  According to these results, the optically-thick (and geometrically-thin) disk is inferred to continue down to close vicinity of the neutron-star surface, and the matter then accretes onto the surface mainly along an equatorial zone to emit the blackbody radiation.
  The blackbody emission from the surface is moderately Comptonized, with a $y$-parameter of $\sim 0.8$ (subsection \ref{ss:323}), by a sort of atmosphere or corona, where the accretion flow would be rather radiatively ineffective and the internal energy would become higher therein.  Like in BHBs, the coronal electron temperature, $T_\textmd{\scriptsize{e}}$, is considered to be determined via a balance between heating by ions and Compton cooling by the {\tt bbody} photons.

To re-examine the very flat hard-tail component claimed by RMD11 using the same data set, we further included five data points in the 31--70 keV range  (subsection \ref{sss:314}).
They indeed exhibit excess above the {\tt diskbb}+{\tt nthcomp}({\tt bbody}) fit (figure \ref{fig:HTs}),
and this effect can be accounted for by a hard PL (with $\Gamma \sim 0$ fixed) with normalization consistent with what was reported by RMD11.
However,  the PL component has a rather marginal significance (confidence level of 93\%), 
when properly considering systematic errors associated with the HXD-PIN NXB subtraction.
In addition, the suggested phenomenon, with $\Gamma \sim 0$, is rather unusual.
For these reasons, we refrain from further discussion on this issue.

  According to the virial theorem, matter in a standard accretion disk must radiate half its gravitational energy which is released by the time it reaches $R_\textmd{\scriptsize{in}}$.  Assuming that the corona is localized in a region around the neutron star, the remaining half energy, plus the energy released from $R_\textmd{\scriptsize{in}}$ to the neutron-star radius $R_\textmd{\scriptsize{NS}}$, will power the blackbody from the neutron-star surface and the corona.  Therefore, when spherically integrated, the disk luminosity is expected to be comparable to that of the Comptonized blackbody, assuming $R_\textmd{\scriptsize{in}}\sim R_\textmd{\scriptsize{NS}}$.  However, the observed fluxes of {\tt diskBB} and {\tt nthcomp} ({\tt bbody}), denoted $F_\textmd{\scriptsize{disk}}$ and $F_\textmd{\scriptsize{bb}}$ respectively, will depend on our viewing angle in different ways.
  As noticed above, the blackbody-radiating area would have a shape of a thin and short cylinder along the equatorial region of the neutron star, and so would be the corona.  Then we can estimate the inclination angle $\theta_\textmd{\scriptsize{i}}$ of the disk as
\begin{eqnarray}
\tan\theta_\textmd{\scriptsize{i}} &=& \frac{\pi}{2}\frac{F_\textmd{\scriptsize{bb}}}{F_\textmd{\scriptsize{disk}}}
\label{eq:inclination}
\end{eqnarray}
\citep{Mitsuda1984}.  The observed ratio of $F_\textmd{\scriptsize{bb}}$ (in 0.1--100 keV) to $F_\textmd{\scriptsize{disk}}$ (in 0.1--50 keV) is $1 : 2.3$, which constrains the inclination angle as $\sim 34^\circ$.  This value is not so different from the value of $45^\circ$, which we tentatively assumed in the data analysis.
  
  The successful fit incorporated two absorption edges that are considered of celestial origin.  Their energies turned out to be $7.1_{-0.1}^{+0.3}$ keV and $8.5\pm 0.4$ keV, corresponding to neutral Fe-K edge and that of highly ionized Fe-K, respectively.  However, the data lack associated Fe-K fluorescence lines  (with an upper-limit equivalent width of 10 eV assuming them to be narrow), like in a previous study \citep{Dai2006}; the absence of lines might be attributed to the geometry of the absorber.

\subsection{Geometry in the Hard state}\label{ss:43}
The hard-state spectra have been explained successfully by either a {\tt diskBB}+{\tt compPS}({\tt bbody}) model or a {\tt bbody}+{\tt  compPS}({\tt diskBB}) model.
  Based on the fit goodness alone, we cannot tell which of the two model forms is more appropriate (subsection \ref{ss:323}).
  Let us then examine the two solutions for their physical appropriateness.  In the {\tt diskBB}+{\tt  compPS}({\tt bbody}) modeling, the derived radii are $R_\textmd{\scriptsize{bb}} = 10\pm 2$ km and $R_\textmd{\scriptsize{in}} = 21\pm 4$ km.  The former is nearly the same as that of a neutron star, and the latter is larger than that; this is physically reasonable.  In the {\tt bbody}+{\tt compPS}({\tt diskBB}) modeling, in contrast, the radii are $R_\textmd{\scriptsize{bb}} = 41\pm 8$ km and $R_\textmd{\scriptsize{in}} = 4.2\pm 0.6$ km, which are not physical since these would mean that the inner edge of the disk is smaller than the neutron star radius.  Thus, the {\tt diskBB}+{\tt compPS}({\tt bbody}) model, which has the same form as our final soft-state model, is clearly more favored.

The same October 9 data, 
together with the other two hard-state data sets  (2 and 4 in figure \ref{fig:HNR}), were already analyzed by RMD11.
They used several models which are qualitatively similar to ours,
each consisting of a softer optically-thick emission and a harder Comptonized component.
Specifically, their M1a model has a form of {\tt diskBB}+{\tt nthcomp}({\tt diskBB}), and their M1b model is described as {\tt diskBB}+{\tt nthcomp}({\tt bbody}).
  Although their M1b model is essentially the same as our final solution, RMD11 concluded M1a to be the best model.
Furthermore, presumably due to differences in the employed Compton codes and in the Fe-K line modeling, 
our results and theirs show some discrepancies as listed below.
\begin{enumerate}
%
\item 
Our successful final fit with {\tt diskBB+compPS(bbody)} to the hard-state data has given 
a reasonable inner disk radius as $R_\textmd{\scriptsize{in}}= 21\pm 4$ km (table 3),
while RMD11 found, with their favourite M1a model and assuming the same distance,
too small a value of $R_\textmd{\scriptsize{in}}= 6.7\pm0.2$ km.
%
\item 
The value of $T_\textmd{\scriptsize{in}} \sim 0.3$ keV we measured 
in the hard state is lower than that ($\sim 0.7$ keV) in the soft state,
while RMD11 reported (again with the M1a model) opposite behavior,
which would be inconsistent with the change in the mass accretion rate.
%
\item 
The values of $N_\textmd{\scriptsize{H}}$ 
which we obtained are consistent with being the same between the two states, 
while this does not apply to the M1a results of RMD11.

\item
Even if we compare our results with the M1b fit by RMD11
(which has the same model form as ours but is not favored by RMD11),
they obtained too large a value of $R_\textmd{\scriptsize{bb}}=18 \pm 5$ km,
while our estimate  of $R_\textmd{\scriptsize{bb}} = 10 \pm 2$ km (table 3)
is fully consistent with the neutron-star radius.
Furthermore, we found  in both states $T_\textmd{\scriptsize{in}}< T_\textmd{\scriptsize{bb}}$,
while this is not necessarily the case in the M1b results of RMD11 as they admit. 
(No information as to these points is available with their M1a modeling.)
\end{enumerate}
Considering these, 
as well as the appropriateness of {\tt compPS} to very hard spectra,
we consider that our hard-state results provide
a better description of the October 9 data.

Although our final spectral model for the hard state has the same composition as that in the soft state (except the edges and line), the model parameters are considerably different, as we see below.
  The standard disk is truncated at a larger radius ($\sim$20 km) than in the soft state. 
  Inside that radius, the flow becomes optically thin and geometrically thick.  This optically-thin disk itself would be identified with the Compton corona, with a much larger $y$-parameter of $\sim 2$ than in the soft state ($y\sim 0.8$).  Unlike that of the soft state, the accretion flow around the neutron star is considered to be rather spherical, because nearly the whole neutron-star surface is inferred to be emitting the blackbody seed photons, which are then inversely-Comptonized by the corona.  Although it is still possible that the standard disk is partly covered with the corona and part of its emission is Comptonized, the Comptonization of the blackbody should be rather dominant because the reflection solid angle is not so large ($\Omega/2\pi \sim 0.6$).
 
 The observed flux ratio of {\tt compPS}({\tt bbody}) to {\tt diskBB} is about $6 : 1$.  Thus, the relative dominance of the two components has reversed as compared to the soft state.  This is explained, at least qualitatively, by the following two effects.  One is the increase in $R_\textmd{\scriptsize{in}}$, which would reduce $F_\textmd{\scriptsize{disk}}$ (as $\propto R_\textmd{\scriptsize{in}}^{-1}$ if the total luminosity is conserved) and enhance $F_\textmd{\scriptsize{bb}}$.  The other is the geometrical change in the corona and the blackbody, from the short cylinder in the soft state to the nearly spherical configuration in the hard state.
  
  The fit required a narrow line at 6.4 keV, with an equivalent width of $\sim 22$ eV.  This is interpreted as (nearly) neutral Fe-K line, produced when the cool accretion disk is irradiated by the very hard continuum.

  In the picture as derived so far, the motion of the accreting matter is expected to change from Keplerian to nearly free-fall like, at a radius of $R_\textmd{\scriptsize{in}} \sim R_\textmd{\scriptsize{c}}$, where $R_\textmd{\scriptsize{c}}$ is a typical radius of the corona.  Therefore, the infall speed of the matter in the corona may be written as
\begin{eqnarray}
v(r) &=& g \sqrt{ \frac{2GM_\textmd{\tiny{NS}}}{r} } \qquad (0 < g < 1),
\label{eq:ffv}
\end{eqnarray}
where $r$ is the distance from the neutron star, $G$ is the gravitational constant, $M_\textmd{\tiny{NS}}$ is the neutron star mass, and $g$ is a numerical factor of order unity.  Meanwhile, the observed total luminosity is represented as 
\begin{eqnarray}
L_\textmd{\scriptsize{obs}} &=& f  \frac{GM_\textmd{\tiny{NS}} \dot{M}}{R_\textmd{\tiny{NS}}}  \qquad (0 < f < 1),
\label{eq:Lobs}
\end{eqnarray}
where $\dot{M}$ is the mass accretion rate, and the numerical factor $f$ becomes less than 1 when there are outflows or when a fraction of the released energy is stored as internal energies of the neutron star.  This $\dot{M}$ is also related with the radius and electron density $n_\textmd{\scriptsize{e}} (r)$ as
\begin{eqnarray}
\dot{M} &=& 4 \pi r^2 v(r) m_\textmd{\scriptsize{p}} n_\textmd{\scriptsize{e}}(r),
\label{eq:corona_ne}
\end{eqnarray}
where $m_\textmd{\scriptsize{p}}$ is the proton mass.  Using the three equations, and performing integration from $r = R_\textmd{\tiny{NS}}( = $10 km) to $r=R_\textmd{\scriptsize{c}}$, the optical depth of the corona is represented as 
\begin{eqnarray}
\tau &=& \sigma_\textmd{\tiny{Th}} \int_{R_\textmd{\tiny{NS}}}^{R_\textmd{\tiny{c}}} n_\textmd{\scriptsize{e}}(r) dr \nonumber\\
&=&  0.04 \times \left\{ 1-\left(\frac{R_\textmd{\scriptsize{c}}}{10  \textmd{km}} \right)^{-\frac{1}{2}} \right\} \cdot \frac{1}{fg}\; ,
\label{eq:corona_tau}
\end{eqnarray}
where $\sigma_\textmd{\tiny{Th}}$ is the Thomson cross section.  So, if we assume $R_\textmd{\scriptsize{in}} \sim R_\textmd{\scriptsize{c}}\sim 20$ km and the product $fg$ takes a value of $\sim 0.01$, we can explain the observed value of $\tau$.

\section{Summary}
Utilizing the archival Suzaku data of Aquila X-1 in an outburst, we successfully obtained high-quality spectra on two occasions, and performed quantitative model fitting to both spectra.  The obtained results are summarized as follows.
\begin{enumerate}
\item The 0.8--31 keV spectra obtained on 2007 September 28 exhibited typical characteristics of the soft state, with the 2--20 keV luminosity being $\sim 8\%$ of the Eddington limit.  In contrast, the 0.8--100 keV spectra on 2007 October 9, much harder as usually seen in the hard state, had a 2--100 keV luminosity which is $\sim 1\%$ of the Eddington limit.

\item The results of the soft state analyses are generally consistent with those from previous studies, including RMD11.  There are two emission components; a disk blackbody emission from a standard disk, and a weakly Comptonized blackbody emission arising from the neutron-star surface.

\item The flat hard-tail component, claimed by RMD11 based on the same Suzaku observation on September 28, becomes rather inconclusive when considering the systematic background errors.

\item The hard-state spectra can be expressed by the sum of a disk blackbody and a Comptonized blackbody.  The Compton seed photons are provided by a blackbody emission from the neutron star surface.

\item In the hard state, a standard accretion disk is inferred to be truncated at $\sim 20$ km, at which point the accretion flow turns into an optically-thin nearly spherical flow.  The blackbody emission from the neutron-star surface is strongly Comptonized with a $y$-parameter of $\sim 2$.  For these two reasons,
the overall spectrum is much harder than that in the soft state, even though the basic emission components are the same.
\end{enumerate}

\end{document}